\documentclass[12pt,english,preprint,authoryear]{elsarticle}
\usepackage[T1]{fontenc}
\usepackage[latin9]{inputenc}
\usepackage{float}
\usepackage[margin=3cm]{geometry}
\usepackage{amsmath}
\usepackage{graphicx}
\usepackage{setspace}
\usepackage{amssymb}
\usepackage{esint}

\let\latexo\o

\usepackage{amsfonts}
\usepackage{ae,aecompl}
\usepackage{lineno}
\usepackage{babel}

\journal{Theoretical population biology}

\providecommand{\tabularnewline}{\\}

\makeatother

\begin{document}

\begin{frontmatter}
\title{Indirect effects of primary prey population dynamics on alternative prey}

\author[trom]{Fr{\'e}d{\'e}ric Barraquand\corref{cor1}}
\author[stand,usmc]{Leslie F. New}
\author[abdn]{Stephen Redpath}
\author[stand,glasg]{Jason Matthiopoulos}     

\address[trom]{Department of Arctic and Marine Biology, University of Troms\latexo}
\address[stand]{Centre for Research into Ecological and Environmental Modelling, University of St-Andrews}
\address[usmc]{US Marine Mammal Commission}
\address[abdn]{Institute of Biological and Environmental Sciences, University of Aberdeen}
\address[glasg]{Institute of Biodiversity, Animal Health and Comparative Medicine, University of Glasgow}

\cortext[cor1]{
Corresponding author. E-mail: frederic.barraquand@uit.no\\
Fr{\'e}d{\'e}ric Barraquand, University of Troms\latexo \\
Department of Arctic and Marine Biology, Naturfagbygget\\
Dramsvegen 201, 9037 Troms\latexo, NORWAY\\
Tel: (+47)(+47)77623169
}

\newpage

\begin{abstract}
We develop a theory of generalist predation showing how alternative
prey species are affected by changes in both mean abundance and variability
(coefficient of variation) of their predator's primary prey. The theory is motivated by the
indirect effects of cyclic rodent populations on ground-breeding birds,
and developed through progressive analytic simplifications of an empirically-based model.
It applies nonetheless to many other systems where primary prey have fast life-histories and
can become locally superabundant, which facilitates impact
on alternative prey species. In contrast to classic apparent competition
theory based on symmetric interactions, our results suggest that predator effects on alternative
prey should generally decrease with mean primary prey abundance, and
increase with primary prey variability (low to high CV) 
- unless predators have strong aggregative responses, in which case these results can be reversed.
Approximations of models including predator dynamics (general
numerical response with possible delays) confirm these results but
further suggest that negative temporal correlation between predator
and primary prey is harmful to alternative prey.  We find in general that predator
numerical responses are crucial to predict the response of ecosystems
to changes in key prey species exhibiting outbreaks, and extend the apparent competition/mutualism theory
to asymmetric interactions. 
\end{abstract}


\end{frontmatter}


\section{Introduction}

Most predators switch to alternative prey species when their primary,
favourite prey becomes scarce. The classical prey model of optimal
foraging theory reflects such a switch, showing how the diet breadth
of consumers increases when consumers' favourite food items are at
low density \citep{emlen1966role,schoener1971theory,pulliam1974theory,charnov1976ofa,pyke1977ofs,pyke1984oft,stephens1986ft},
and shrinks again at high favourite food density. For community dynamics,
this implies that alternative prey species are influenced by the dynamics
of the primary (i.e. preferred) prey through their influence on the
predator's functional \citep{oaten1975switching} and numerical \citep[e.g.][]{holt1987short,wilson2001functional}
response to shifts in primary prey abundance. Predator numerical
responses can be aggregative through movements \citep{turchin1997ebm}, or demographic through
reproductive changes; in most cases a mixture of both. There is, however,
an essential tension between the consequences of functional and aggregative/numerical
responses for alternative prey. When the primary prey increases in
numbers, individual predators tend to eat less of alternative prey
species, but predators also tend to be more numerous, which potentially
increases the overall impact of predators on alternative prey populations.
The impact of predators on alternative prey results from a balance
between the numerical and functional components of the predator response,
which determines whether the primary prey has a positive or negative
impact on the alternative prey. Apparent interactions can therefore
take the form of competition or mutualism \citep{abrams1998apparent,bety2002shared,brassil2004prevalence},
or even amensalism versus commensalism, if the primary prey has greater
impact on the secondary prey than the reverse, which often seems to
be the case because of differences in overall biomass and densities
\citep{sinclair2003mammal}. Here, we will consider these asymmetric
interactions, i.e. prey 1 indirectly affects prey 2 but not the other
way around. It seems that such asymmetric subsets of the food web
(trophic modules) are not only quite widespread \citep{stouffer2012evolutionary},
but also important to consider from a functional or conservation perspective
\citep{decesare2010endangered,wittmer2012conservation}. 

Predators generally prefer to feed on fast-reproducing species, r-strategists
that invest heavily on reproduction and less on survival, if only
because these prey are easier to catch and have large maximal
densities. Such fast-reproducing species are strongly influenced by
environmental variability, and therefore tend to have dynamics that
are both highly variable \citep{saether2002demographic,sinclair2003mammal}
and often nonstationary \citep[e.g.][]{angerbjorn2001geographical}.
The influence of variability and nonstationarity of primary prey dynamics
on alternative predation is not yet well developed in the otherwise abundant apparent
competition/mutualism literature \citep{holt1994ecological,abrams1998apparent,brassil2004prevalence,brassil2006can,schmidt2008numerical},
despite its relevance to a number of species of scientific and conservation
importance (see \citealp{schmidt2008numerical} for examples). In
this paper we investigate how the interplay between the mean and variability
of primary prey abundance affect alternative prey species demography. 
We restrict the definition of alternative prey in two ways in this article. 
First, it is a species that - unlike the primary prey - is either not available 
or not nutritious enough for the predator to specialize on it year-round. 
Second, which might be a corollary of the first, the alternative prey cannot
drive the aggregative/numerical response of the predator, which is mostly influenced by the primary prey \citep[e.g.][]{new2011hen,new2011modelling}.  
These restrictions make the asymmetric nature of the interaction
all the more likely, and correspond well to the examples described below.

A typical terrestrial example of highly variable primary prey are rodents
such as voles and lemmings (subfamily \emph{Arvicolinae}), that exhibit
large-amplitude cycles, especially in boreal and arctic regions. These
cycles are often non-stationary, because rodent vital rates react
to trends in climatic variables \citep{kausrud2008linking}. As \citet{lack1946competition}
remarked early on, rodents are preyed upon by an important guild of
avian and mammalian predators. This guild includes foxes and mustelids,
on the mammalian side, and on the avian side, raptors and some other
birds such as skuas and corvids. All these rodent-eating predators
have the habit of switching to alternative food sources when the rodent
cycle is at a trough; often such alternative prey consists of eggs
and juveniles of gamebirds, waterfowl, etc. \citep[see][for more details]{valkama2005birds}.
Thus, the breeding success of many bird species is severely impacted
by the population dynamics of rodents \citep{lack1946competition,summers1998breeding,wilson2001functional,blomqvist2002indirect,bety2002shared,valkama2005birds,schmidt2008numerical}.
It has even been hypothetized that bird breeding habitat in the Arctic
is partly determined by association with cylic rodents \citep{blomqvist2002indirect,gilg2010explaining}. 

Taking a larger view of the food web, such variable primary prey are
often key species in their ecosystems, with a large number of links
to other species \citep{sinclair2003mammal,jordan2009keystone,krebs2011lemmings}.
Examples of key prey whose dynamics can influence alternative prey
species range from lemmings and hares (\emph{Lepus} sp.) in boreal
and arctic landscapes \citep{krebs2011lemmings} to wildebeest (\emph{Connochaetes}
sp.) in the Serengeti \citep{sinclair2003mammal}; or in marine ecosystems,
from sandeels (family \emph{Ammodytidae}; \citealp{matthiopoulos2008getting})
to small, overabundant pelagic fishes that create so-called ``wasp-waist''
ecosystems \citep{cury2000small}. \textbf{ }Though such species are
sometimes referred to as ``keystone'' \citep[e.g.][]{cornulier2013europe},
this is a misnomer, since their importance is largely due to their
large maximal biomass at peak densities (unlike keystone species which
have a disproportionate effect on the ecosystem per unit of biomass,
\citealp{power1996challenges}). Hence, we refer to rodents and their
r-strategist counterparts in other ecosystems simply as ``key''
prey species.

In a previous paper, \citet{matthiopoulos2007sensitivity} developed
a model for predation by harriers (\emph{Circus cyaneus}) on red grouse
chicks (\emph{Lagopus l. scoticus}, an alternative prey to voles \emph{Microtus
agrestis}) in Scottish moors, for different levels of abundance of
voles and pipits (\emph{Anthus pratensis}), both of which are important
in the harriers' diet. This was done using empirically measured multispecies
functional and aggregative responses. The study was largely motivated
by management of a human - wildlife conflict; hunters typically want
to shoot more grouse and see less harriers, while conservationists
wish to protect the hen harrier, an endangered raptor in the UK \citep{thirgood2008hen}.
The impact of harriers on grouse is managed through a number of techniques,
that range from the illegal killing of raptors to diversionary feeding
\citep{redpath2001does}. A modelling assessment of the various management
techniques is given in \citet{new2011modelling}. We use this system
as a key empirical example to motivate general theory rather than
the object of study per se. Red grouse corresponds well to our abovementioned definition
of alternative prey. 

Field vole numbers - the primary prey of harriers - were assumed for
simplicity to be a constant in \citet{matthiopoulos2007sensitivity}.
However, vole abundance can vary greatly from year to year, and harriers
react numerically to these variations \citep{redpath2002field}. The
model we develop here relaxes the assumption of constant vole primary
prey availability. The way we represent primary prey dynamics is akin
to a resource pulse \citep{holt2008theoretical,schmidt2008numerical}.
Through a series of progressive approximations of the detailed, empirically-based
model of vole-harrier-grouse dynamics, we formulate a simplified model
that can apply to a number of other prey communities that share the
same basic features with the rodent-raptor-gamebird system, which
are: (1) strongly fluctuating primary prey and (2) asymmetric interactions.

We initially extend the vole-harrier-grouse model to account for vole
variability, and show how grouse average numbers critically depend
on the temporal average of grouse breeding success and therefore,
temporal average of the total response (total number of grouse chicks
killed = aggregative $\times$ functional response). This then leads
us to a more detailed examination of how the average total response
(number of grouse chicks killed) depends on primary prey variability
(i.e. coefficient of variation), and how the effect of primary prey
variability interacts with that of mean primary prey density. As previously
remarked by \citet{schmidt2008numerical}, variability in primary
prey density can affect a predator's average consumption, because
their total response is a nonlinear function of primary prey abundance
(a consequence of Jensen's inequality). We show that the direction
of such effects is driven by mean primary prey abundance. We first
consider the case of fast aggregative responses to primary prey density
(section 2 and 3), but also extend some of these results to delayed
numerical responses (section 4).

\section{The Vole-Hen Harrier-Grouse model}

\subsection{The model}

The model we present here is derived from \citet{matthiopoulos2007sensitivity},
with the addition of fluctuating dynamics for voles. In \citet{matthiopoulos2007sensitivity},
a third prey, pipits, was considered. However, for
simplicitly, in most of this paper, we fix the density of the third
prey, pipits, to a constant. This assumption (relaxed in 
\ref{ap:Dealing}) is justified because pipit numbers vary usually less than vole numbers. 

\subsection{Building blocks}

\subsubsection{The vole model}

Throughout the manuscript, we use the letters $V$ to denote vole
or more generally primary prey density, and $G$ for grouse or alternative
prey. We use two alternative descriptions of vole population dynamics,
a deterministic model producing limit cycles (based on \citealp{maynardsmith1973stability})
or a simple probability distribution of values (either lognormal or
gamma, because abundance distributions are non-negative and skewed
to the left).\textbf{ }These contrasting formulations are justified
biologically because observed temporal fluctuations in voles range
from deterministic-looking cycles to lognormal noise driven by abiotic
variables \citep{stenseth1999population}, and mathematically because
the periodicity of vole dynamics does not influence the main results
which are driven by the mean and coefficient of variation. 

The Maynard-Smith and Slatkin (MSS) model is given by eq. \ref{eq:mss},
in which $R_{V}$ is the maximal vole growth rate, $K_{V}$ is the
threshold density marking the onset of density-dependence, \textbf{$k$
}is the time lag of density-dependence (one or two timesteps), and
$\gamma$ is an exponent characterizing the abruptness of density
dependence. 

\begin{equation}
V_{t+1}=\frac{R_{V}V_{t}}{1+(V_{t-k}/K_{V})^{\gamma}}\label{eq:mss}\end{equation}
Alternative dynamical models such as autoregressive models on a linear
scale \citep{royama1992analytical} provide similar results for the
predation models considered below, so long as we consider aggregative predator responses.

\subsubsection{The grouse model and harrier predation}

The grouse model is based on \citet{matthiopoulos2007sensitivity},
and is essentially a delayed version of the Hassell model for population
cycles, including some predation on the maximal birth rate (eq. \ref{eq:grouse_model}
below)

\begin{equation}
G_{t+1}=\frac{(B(G_{t},V_{t})+S)G_{t}}{(1+G_{t-1}/K_{G})^{\beta}}\label{eq:grouse_model}\end{equation}

Here, $S$ is the grouse survival rate, $K_{G}$ is a threshold for
density-dependence, and $\beta$ controls the shape of density-dependence.
$B(G_{t},V_{t})$ is the birth rate, modelled as a function of both
prey densities \begin{equation}
B(G_{t},V_{t})=\max(0,\, B_{0}-H_{t}f(B_{0}G_{t},V_{t})/G_{t})\end{equation}
 $H_{t}$ is harrier density and $f(B_{0}G_{t},V_{t})$ is the raptor's
functional response (equation 5) to the densities of grouse chicks
and voles (see Table 1). Assumptions about $H_{t}$ are either $H_{t}=H_{0}$
(a constant) or $H_{t}=\alpha_{1}V_{t}+\alpha_{2}P$ where $P$ is
pipit density and $\alpha_{1},\alpha_{2}$ relate vole and pipit density
to $H_{t}$ (this is a linear aggregative response, as in \citealp{matthiopoulos2007sensitivity}).
Later, in section 4, we consider population dynamics for $H_{t}$. 

Note that we can also write the model as

\begin{equation}
G_{t+1}=\frac{\textrm{max}(B_{0}G_{t}-H_{t}f(B_{0}G_{t},V_{t})+SG_{t},SG_{t})}{(1+G_{t-1}/K_{G})^{\beta}}\end{equation}

\subsubsection{Harrier functional response}

The functional response relates the intake rate of harrier pairs to
the numbers of voles, grouse young, and pipits (considered constant
here). Assuming the density of the $i^{th}$ prey species is denoted
by $N_{i}$, the functional response on species $i$ can be written

\begin{equation}
f_{i}=\frac{a_{i}N_{i}^{m_{i}}}{1+\sum_{j}a_{j}h_{j}N_{j}^{m_{j}}}\end{equation}
where parameters $a_{i},m_{i},h_{i}$ have been estimated empirically
in \citet{smout2010functional}, with values given in Table 1 (for
consistency with \citealp{smout2010functional}, $i=1$ for grouse,
2 for pipit, 3 for vole). We use this specific parametrization of
the model so as to consider a biologically relevant scenario (illustrated
in Fig. 1); more general functional forms are considered later on. 

\begin{center}
\begin{figure}[H]
\begin{centering}
\includegraphics[width=12cm]{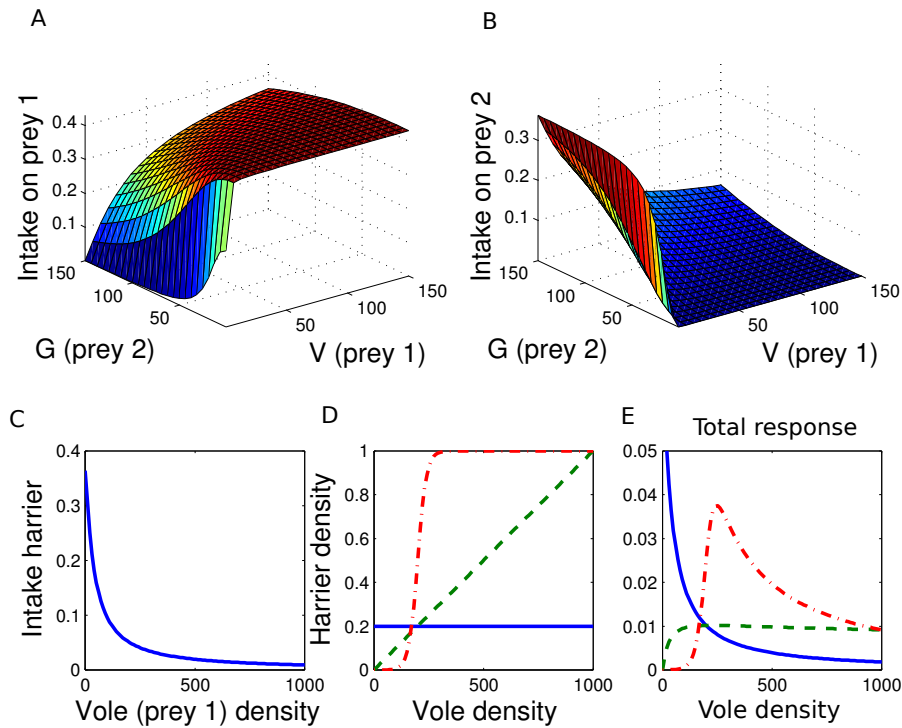}
\par\end{centering}

\caption{The functional response of harriers.\textbf{ }A and B show the multispecies
functional response of harriers (pipit density set to
zero) on voles (A) and grouse (B), while in C we see the functional
response on grouse (prey 2) given vole density (prey 1). Three kinds
of aggregative responses (D) - constant (plain line), linear (dashed)
and sigmoid (dash-dotted) - eventually lead to three kinds of total
response in E (total intake on prey 2 as a function of prey 1= aggregative
response to prey 1 $\times$ functional response on prey 2 as a function
of prey 1). All functional response parameters are identical to those
estimated in Smout et al. (2010). }

\end{figure}

\par\end{center}

\begin{table}
\caption{\textbf{Definition and values of parameters for the harrier predation
model. }}

\begin{centering}
\begin{tabular}{cccc}
\\
\hline 
\textbf{Parameter name} & \textbf{Symbol} & \textbf{Baseline value} & \textbf{Unit}\tabularnewline
\hline
Max vole population growth rate & $R_{V}$ & 40 & NA\tabularnewline
Threshold for density-dependence in vole & $K_{V}$ & 100 & .$km^{-2}$\tabularnewline
Exponent for density-dependence in voles & $\gamma$ & 4.5 & NA\tabularnewline
Grouse baseline survival rate & $S$ & 0.5 & NA\tabularnewline
Threshold for density-dependence in grouse & $K_{G}$ & 285 & $km^{-2}$\tabularnewline
Maximal fertility in grouse & $B_{0}$ & 2.61 & NA\tabularnewline
Exponent for density-dependence in grouse & $\beta$ & 6 & NA\tabularnewline
Discovery rate of vole in harrier FR & $a_{1}$ & 3.78 & NA\tabularnewline
Discovery rate of grouse in harrier FR & $a_{2}$ &  0.000673 & NA\tabularnewline
Discovery rate of pipit in harrier FR & $a_{3}$ & 1.90 & NA\tabularnewline
Handling time of vole in harrier FR & $h_{1}$ & 2.32 & NA\tabularnewline
Handling time of grouse in harrier FR & $h_{2}$ & 2.74 & NA\tabularnewline
Handling time of pipit in harrier FR & $h_{3}$ & 1.67 & NA\tabularnewline
Exponent in functional response vole & $m_{1}$ & 1.14  & NA\tabularnewline
Exponent in functional response grouse & $m_{2}$ & 2.51 & NA\tabularnewline
Exponent in functional response pipit & $m_{3}$ & 1.18 & NA\tabularnewline
Time spent foraging by a harrier each year & $q$ & 900 & $h$\tabularnewline
Coefficient for vole in numerical response & $\alpha_{1}$ & 0.000124 & NA\tabularnewline
Coefficient for pipit in numerical response & $\alpha_{2}$ & 0.00263 & NA\tabularnewline
\hline
\\
\end{tabular}
\par
\end{centering}

\textbf{Note:} FR= functional response. 
\end{table}

\subsection{Effect of vole variability on the full community model}

Vole population dynamics (Fig. 2) modify grouse population cycles
through harrier predation, though not to a level generating synchrony
between vole and grouse for realistic harrier densities (see Discussion).
Using simulations of the MSS model to represent vole dynamics (section
2.2.1), as in Fig. 2, as well as more simplifed probabilistic representations
(lognormal noise), we found that the temporal average of grouse population
size, the temporal average of per capita grouse breeding productivity,
and the temporal average of the total response are all strongly correlated
(Fig. 3), and respond similarly to variations in mean and variability
(i.e. coefficient of variation) in vole density. In \ref{ap:Correlations}, we show why such correlations are expected in models for regulated
(albeit fluctuating) populations. Therefore, in the following section,
we focus only on the temporal average of the total response, which
is more general and easier to compute. 

\begin{figure}[H]
\begin{centering}
\includegraphics[width=10cm]{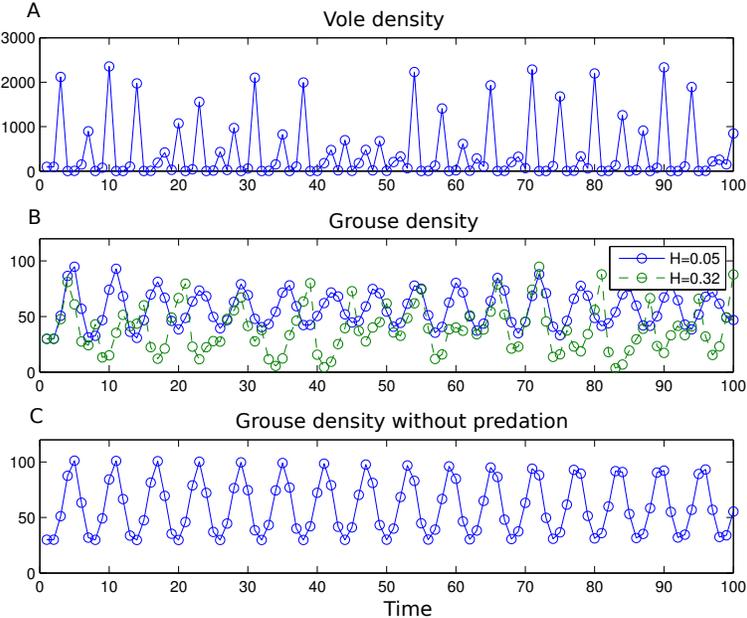}
\par\end{centering}

\caption{Population dynamics of grouse with (B) or without (C) predation on
grouse, with a variable vole dynamics (A). No predator numerical/aggregative
response were considered here, only constant numbers of harriers.
In plate B, the blue plain line corresponds to $H=0.05$, and the
green dashed line to $H=0.32$ (maximum observed harrier densities
in Scotland). See Table 1 for other parameters. }

\end{figure}

\begin{figure}[H]
\begin{centering}
\includegraphics[width=14cm]{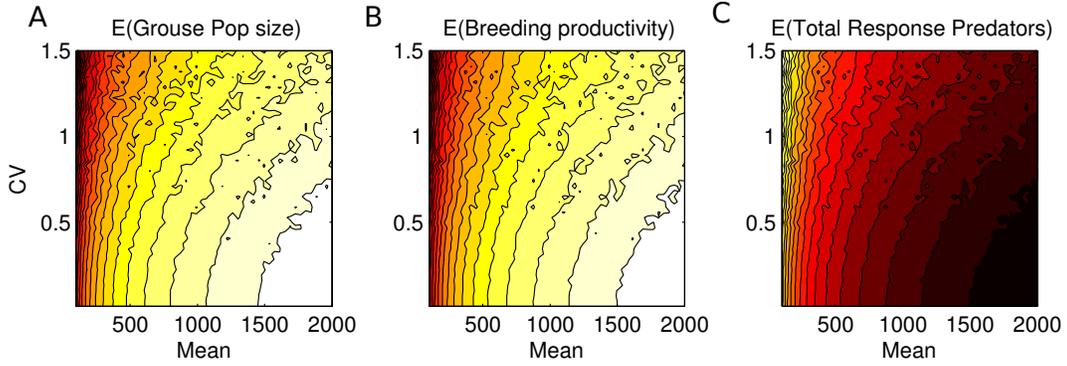}
\par\end{centering}

\caption{Congruence of effect of mean vole density and variability (coefficient
of variation, CV) on average grouse population size (A), average breeding
success (B), and average number of chicks predated (C, i.e. average
total response).\textbf{ }Vole population dynamics is given by a simple
lognormal probability distribution, whose mean and CV are varied in
the x and y axes. Parameters are otherwise equal to those in Table
1 (with empirically measured aggregative response instead of constant
predator numbers).}

\end{figure}

\section{Average total response on alternative prey: effect of mean density
and variability in primary prey}

In the preceding section, we showed how the total response, averaged
over the years, provides a good indicator of predator impact on secondary
prey,\textbf{ }when the predator aggregates to the primary prey. This
will be the case in situations where predators mostly impact the juvenile
component of the alternative prey population (see \ref{ap:Correlations}). 
Here, we therefore analyze the temporal average $\bar{T}$ of
the total response $T(V_{t})$, \begin{equation}
\bar{T}=\frac{1}{n}\sum_{t=1}^{n}T(V_{t})=\frac{1}{n}\sum_{t=1}^{n}f(V_{t})H(V_{t})\label{eq:temp_av_tot_resp}\end{equation}

In the long run, this value converges to its expectation over realizations,
given a probability distribution $\rho(v)$ for voles densities. Because
of the ergodic theorem, this is true irrespective of possible autocorrelation
in the $V_{i}$ values and leads to \begin{equation}
\lim_{n\rightarrow\infty}\frac{1}{n}\sum_{t=1}^{n}T(V_{t})=\mathbb{E}(T(V))=\int_{0}^{\infty}\rho(v)T(v)dv\label{eq:expected_tot_resp}\end{equation}

\subsection{General formulation using gamma functions}

Using Taylor approximations, it is possible to derive insights on
how the average total response is affected by the mean and variability
of primary prey density (this works best for $\textrm{CV}<0.3$).
However, because the primary prey has usually very variable dynamics
(i.e. CV $\geq$ 1), it is worthwhile to derive results for {}``large
noise'' conditions. Using relatively mild assumptions for the functional
and aggregative responses, it is possible to derive such results.
A typical form for the functional response on prey 1 as a function
of prey 2 could be $f(V)=f_{0}e^{-\lambda V}$ (grouse chicks killed
as a function of rodent density $V$), and for the aggregative response,
some power function $H(V)=\alpha V^{l}$ (number of predators purely
driven by the primary prey density). More complicated functions are
possible, but do not change the main argument. In general, the functional
response $f$ depends on both $G$ and $V$; the reason why we approximate
$f(G,V)$ by $f(V)$ is because of the stronger dependency on $V$
(Fig. 1), especially when the grouse population is regulated. 
Similar functions are presented in \citet{schmidt2008numerical}, 
and the \ref{ap:foraging} describes how a nearly exponential decrease of the functional response
can be obtained even in cases of optimal predator foraging, where we would usually expect a function with both concave and convex parts. 

This combination of functional forms yields a total response $T(V)$
of the kind $A_{l}V^{l}e^{-\lambda V}$ (Fig. 4A, with the exponent
$l$ changing how the total response accelerates at low densities,
and a total response maximum at $l/\lambda$). The total response,
for efficient predators that forage optimally, might drop faster than
exponentially, e.g. steeper declines in the functional response, which
complicates the mathematics but does not change the qualitative results.
Because it will be important to the computations, let us mention that
$AV^{l}e^{-\lambda V}$ has the same functional form as the probability
distribution of a gamma random variable (presented in \ref{ap:Sensitivity}). 
Then, using a gamma random variable to model the probability distribution
of primary prey values, we reduce the problem of computing the average
total response to computing particular integrals of gamma functions
(\ref{ap:Sensitivity}). 

We found that in the case $l=1$ (linear aggregative response $H(V)=\alpha V$),
increasing the mean rodent density $m$ first increases predation
on prey 2, up to a maximum after which it starts decreasing (\ref{ap:Sensitivity} and Fig. 4). 

This means that the deterministic total response curve $T(V)$ is
a good indicator of the effect of the average rodent density (denoted
here $m$) on $E(T(V))$. The relationship may seem intuitive - the
effect of the mean on the average total response is the same as that
of that of the deterministic vole density on the deterministic total
response - but verification is important for two reasons. First, there
is substantial variability around the mean (e.g. $\textrm{CV}=1$,
see Fig. 4), and second, the variability is asymmetrically distributed
(due to the gamma distribution, which corresponds well to the shape
of real abundance values). 

In contrast to the case with linear aggregative response, when $l=0$
(contant predator numbers, $H(V)=H_{0}$), we have a constant decrease
of average predation $\bar{T}$ with mean rodent density $m$, again
predictable from the deterministic $T(V)$ curve. 

The effect of variability is not so intuitive, though the curvature
of the total response gives some indication. For $l=1$ ($H(V)=\alpha V$),
increasing variability slightly increases predation for low primary
prey mean $m,$ then decreases predation for intermediate $m$ values,
and for very large $m$, increasing CV increases $\bar{T}$
(\ref{ap:Sensitivity} for a general proof, similar results for $l=5$ in Fig. 4). 
The latter case of large $m$ is unlikely because it would correspond to overabundant key herbivores
all the time. For a large range of mean values $m,$ in the presence
of a strong aggregative response, more primary prey variability will
therefore decrease predation. In contrast, for $l=0$ ($H(V)=H_{0}$),
increased CV always generates increased predation (\ref{ap:Sensitivity}). 
Table 2 summarizes the results. 

\begin{figure}[H]
\begin{centering}
\includegraphics[width=14cm]{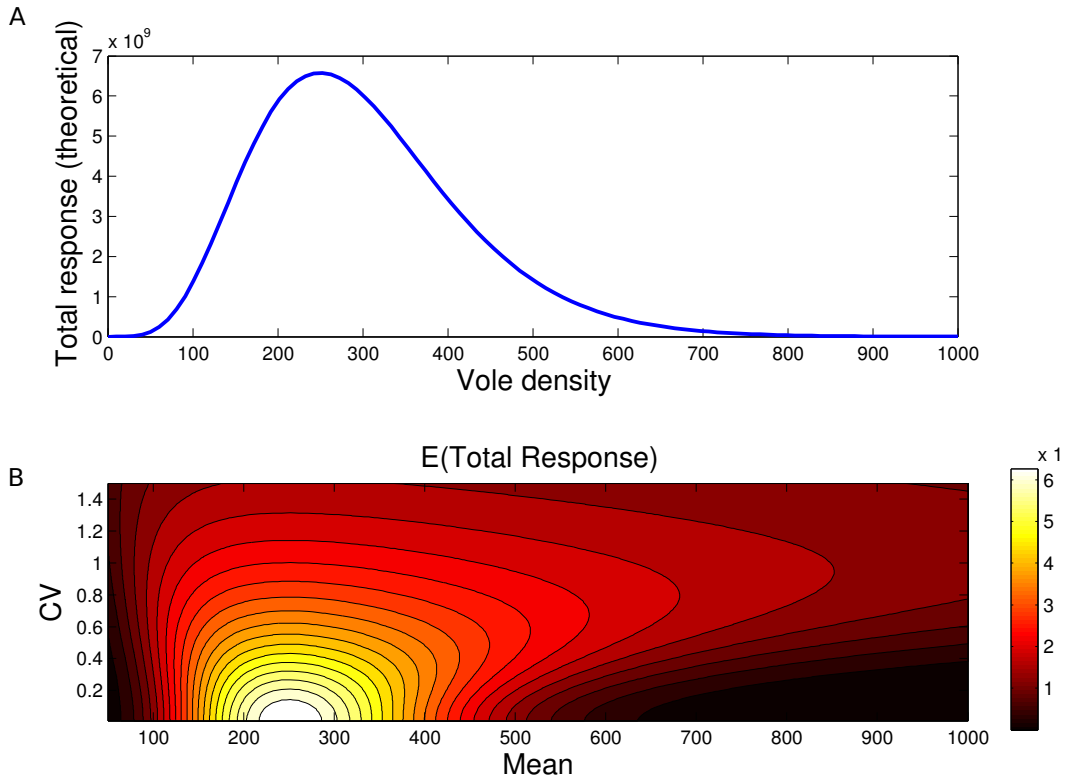}
\par\end{centering}

\caption{Effect of mean and variability of primary prey density on the total
response (B), using a theoretical total response curve $T(V)=A_{l}V^{l}e^{-\lambda V}$
(A). Parameters of $T(V)$: $l=5,\, A=1,\,\lambda=1/50$. For the
Gamma distribution, see values of mean and coefficient of variation
(CV) in the axes, how mean and CV relate to the classic shape and
location parameters is presented in \ref{ap:Sensitivity}. We recover
all the results shown mathematically in the \ref{ap:Sensitivity}. The results
have been obtained through numerical integration of the total response
with $\bar{T}=\int T(v)\rho(v)dv$ where $\rho(v)$ is the pdf of
a Gamma distribution. Numerical simulations with the lognormal distribution
provide similar results. }

\end{figure}

\subsection{Vole-Grouse-Harrier example: numerical integration}

In section 3.1, we used theoretical models without measured parameters.
In this section, we present an example with the empirically measured
parameters for the vole-harrier-grouse system. Fig. 5 shows how increasing
pipit density (which is assumed constant over time) makes the aggregative
response switch from a linear function of vole density to a near-constant,
and modifies the effect of vole variability on grouse. In all cases,
however, it appears that the total response is mostly decreasing and
convex, and therefore increased mean rodent density decreases predation
on grouse while increased rodent density CV increases predation. %
\begin{figure}[H]
\begin{centering}
\includegraphics[width=14cm]{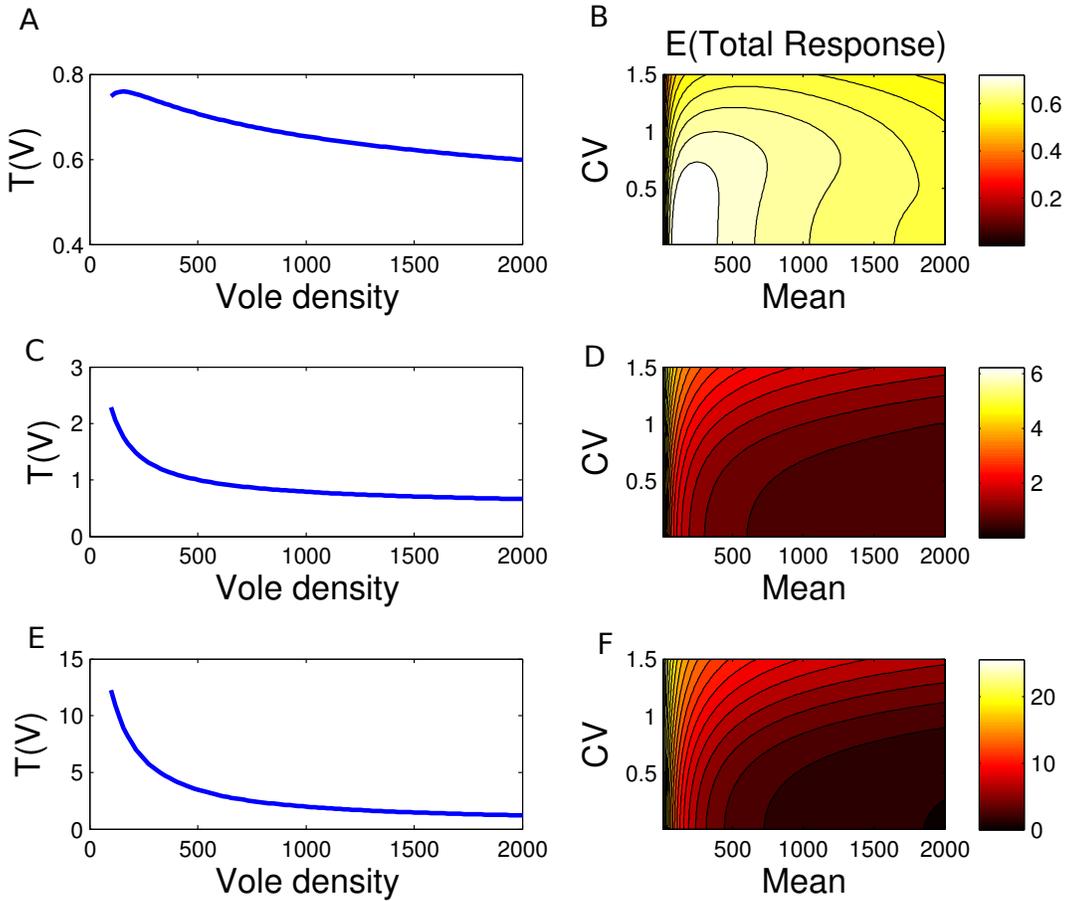}
\par\end{centering}

\caption{Average total response for the vole-harrier-grouse case. A,C,E: The
total response of harriers to varying vole density under different
assumptions of constant pipit density. Top row: $N_{pipit}=0$, Middle
row $N_{pipit}=10$, Bottow row $N_{pipit}=100$ {[}individuals.$km^{-2}${]}.
B,D,F: averaged total response over a Gamma probablity distribution,
computed from numerical integration of the functional and aggregative
responses measured on the vole-harrier-grouse system. Parameters as
in Table 1. }

\end{figure}

\section{Including delays in the predator response: Taylor approximation and
predator-prey covariance}

Delays in the predators' numerical response can arise either in the
aggregative response (due to slow movements) or in the predators demography
(due to the delayed effects of food on reproductive success). Delays
in the aggregative response can happen when the environment is characterized
by strong seasonality, a situation covered extensively in \ref{ap:Seasonality} 
that yields results qualitatively similar to those presented below.
The consequences of a delay in the predators' reproductive response
are difficult to study fully with a parametric model when predation
is important, because of feedbacks between predator and prey populations.
We plan to perform detailed numerical simulations in a future publication.
However, we consider here approximations that take into account the
possible effect of the predator-prey covariance generated by delays
in the predator numerical or aggregative responses, assuming predation
pressure is relatively low, so that prey species 1 and 2 cannot be
considered to be fully synchronized (in which case the average total
response is not correlated to average population size of the alternative
prey). For small predation pressure then, we can focus on the average
total response $\bar{T}$ to find

\begin{equation}
\frac{1}{n}\sum_{i=1}^{n}T(V_{i})=\frac{1}{n}\sum_{i=1}^{n}f(V_{i})H_{i}\label{eq:temp_av_tot_resp_harrier_dyn}\end{equation}

except now the predator density $H_{i}$ has its own dynamics, correlated
to some extent with $V_{i}$. Simplications are obtained by remarking
that

\begin{equation}
E(f(V_{i})H_{i})=\textrm{Cov}(f(V_{i}),H_{i})+E(f(V_{i}))E(H_{i})\label{eq:expected_tot_resp_cov}\end{equation}

We can derive results using classical small-noise approximations to
nonlinear functions. We approximate $E(f(V_{i}))$ by $f(E(V_{i}))+\frac{\sigma_{V}^{2}}{2}f''(E(V_{i}))$
\citep{hilborn1997ecological}. Using the previous formula $f(V)=f_{0}\exp(-\lambda V)$,
and dropping indexes, we obtain

\begin{equation}
f''(\bar{V})=\lambda^{2}f(\bar{V})\Rightarrow E(f(V))\approx f(E(V))(1+\frac{\sigma_{V}^{2}}{2}\lambda^{2}))\end{equation}.

We could also make use of a first order approximation for the functional
response $f(V)\approx f(\bar{V})+f'(\bar{V})(V-V)\approx f(\bar{V})(1-\lambda(V-\bar{V}))$
for small deviations (using interchangeably $f(\bar{V})$ and $f(E(V))$
to simplify notations). The first order approximation helps simplifying
the covariance $\textrm{Cov}(f(V),H)=\textrm{Cov}(f(\bar{V})(1-\lambda(V-\bar{V})),H)=-\lambda f(\bar{V})\textrm{Cov}(V,H)$.
More generally, $\textrm{Cov}(f(V),H)\approx f'(\bar{V})\textrm{Cov}(V,H)$.
This finally leads to the approximate formula for the average number
of grouse killed (=average total response)

\begin{equation}
\bar{T}=E(f(V)H)=\underbrace{-\lambda f(\bar{V})\textrm{Cov}(V,H)}_{\textrm{covariance\,\ effect}}+\underbrace{f(\bar{V})\bar{H}}_{\,\textrm{rodent\,\ mean\,\ effect}}+\underbrace{f(\bar{V})\bar{H}\lambda^{2}\frac{\sigma_{V}^{2}}{2}}_{\textrm{rodent\,\ variance\,\ effect}}\label{eq:general_approx_total_resp}\end{equation}

In this formula, we see 
\begin{itemize}
\item As before, increase in primary prey variability, for a relatively constant
predator number, can increase the predation rate
\item A positive covariance between predator and primary prey density will
decrease the average number of alternative prey killed. A positive
correlation in numbers - as happens in the aggregative response case
- implies that when the individual kill rate on alternative prey is
high (e.g. low rodent density), there are also fewer predators. 
\item In contrast, a negative covariance between predator and primary prey
density does not benefit the alternative prey. This might typically
happen with a delayed predator numerical response to the primary prey
(with some degree of phase opposition in the cycles). 
\end{itemize}
The results above are corroborated by those for delayed aggregative
responses, in \ref{ap:Seasonality}.

\section{Discussion}

We modelled the effect of generalist predation on alternative prey
as a function of the mean and variability of primary prey numbers.
Our work was motivated by an empirically-driven model for the interaction
between voles and grouse through harrier predation, and recents reports
of changes in the mean and sometimes variability of vole or lemming
numbers \citep{ims2008collapsing,schmidt2012response,cornulier2013europe}.
However, other key herbivores (e.g., hares, wildebeest),
with large maximal growth rates compared to other prey species and
large maximal densities, have rather variable and non-stationary dynamics,
and have the capacity to indirectly influence the population dynamics
of alternative prey species. 

We showed that the temporal average of the total response (the total
number of alternative prey killed each year) is a good index of predator
impact for this model. It was indeed strongly correlated with both
alternative prey breeding productivity and average abundance. Our
results apply to a large set of asymmetric apparent interaction modules,
where prey 1 influences prey 2 but not the other way around. 

\begin{table}[H]
\caption{\textbf{Effect of primary prey on alternative prey as a function of
the predator-primary prey covariance.} }
\label{table:results}
\begin{centering}
\begin{tabular}{cccc}
\\
\hline 
\textbf{Sign of covariance} & \textbf{`Worst cycle phase'} & \textbf{$\nearrow$ of E(V) on G} & \textbf{$\nearrow$ of CV on G}\tabularnewline
\hline
Cov(H,V)>0 & intermediate $V$ ({*}) & - at low $V$ then + & + ({*}{*})\tabularnewline
Cov(H,V)$\leq$0 & low $V$ & +  & -\tabularnewline
\hline
\\
\end{tabular}\\

\par\end{centering}

\textbf{Note:} First column: By Cov(H,V) we denote the covariance
(at temporal lag zero) between predator and primary prey. Cov(H,V)>0
corresponds to the aggregative response scenario, Cov(H,V)=0 to constant
or randomly varying predator scenario and Cov(H,V)<0 to a numerical
response scenario (negative correlation is generated by a short demographic
or movement delay). All of our results assume Cov(G,V) $\approx0$
(unsynchronised G and V). Situations with synchronised G and V, corresponding
to high predation pressure on alternative prey, would warrant further
analyses (see \ref{ap:Synchronization}). \\
The three other columns show, from left to right (1) the phase
of the cycle during which the alternative prey suffers the most predation,
i.e. the cycle phase in which management actions would be best implemented
to protect the alternative prey (2) the effect of increasing the temporal
mean of the primary prey density on the alternative prey (3) the effect
of increasing primary prey variability, as measured by the coefficient
of variation (CV). \\
$*$ If the mean primary prey density is very low (very skewed
distribution), intermediate V values from the prey perspective might
not represent intermediate values in the total response of predators.
\\
$**$ Negative effect possible at very large mean primary prey
density, which is unlikely (it would require a permanent overabundance
of key herbivores). 
\end{table}

\subsection{Effect on alternative prey of mean and variability in primary prey
abundance}

The total response on prey 2 as a function of prey 1 density is the
functional response times the aggregative response. Here, the term
functional response differs from its classical use, i.e. it is the
per predator kill rate of prey 2 as a function of the density of prey
1. Such a functional response is derived from the multispecies functional
response, as opposed to Holling's classic single-species functional
responses. The problem of evaluating the effect of primary prey abundance
and variability on alternative predation was simplified by considering
only variation in the primary prey, given it is the prey that dominates
both the functional and aggregative responses. Therefore, in our model,
the total response for prey 2 depended only on prey 1. 

The key quantity in the model was the temporal average of the total
response. The total response is generally a nonlinear function, hence
its temporal average obeys Jensen's inequality, \textbf{$E(T(V))>T(E(V))$
}if \emph{T} is convex, which generates possible effects of prey 1
variability. Usually, approximations of nonlinear functions are used
when such averaging is needed. However, the primary prey (e.g. rodents)
in many real systems exhibits considerable variability in abundance
(i.e. CV $\geq$ 1) which implies that results from linearized models
may have limited applicability. We circumvented this problem using
a more general approach, specifying functional forms that allowed
us to represent primary prey interannual variation as gamma distributions.
Mathematical and computational results using different probability
distributions for primary prey variability agree, and we summarize
these results below and in Table 2.

We found that when predators are in relatively high and constant numbers,
the overall kill rates of alternative prey decline almost exponentially
with respect to primary prey density (such in the vole-harrier-grouse
case when pipits are numerous, Fig. 5). Therefore, predation on alternative
prey decreases when mean primary prey increases, but predation increases
when primary prey variability increases. 

With respect to the effect of the mean, it is an instance of commensalism (0/+) between prey, 
and our finding add  to those of \citet{abrams1998apparent} who found the possibility of mutualism (+/+) in interactive systems. 
With respect to the effect of variability, \citet{schmidt2008numerical}
suggested a similar outcome based on Jensen's inequality, i.e. $E(T(V))>T(E(V))$
because $T$ is convex. In that case, primary prey variability is inherently detrimental to
alternative prey survival. 

However, the literature shows cases where predators can be sparse or even absent when primary prey is low \citep{gilg2006functional}.
In these cases, our theoretical modeling shows that such strong predator
aggregative responses may generate the opposite pattern: 1) Increases
in mean primary prey are detrimental to secondary prey but 2) temporal
variability in primary prey can decrease predation on secondary prey.
The occurrence of such cases might be restricted to high-arctic latitudes
however, and calls for more empirical research. 

More generally, our results suggest that empirical measurements of
predators' aggregative responses to primary prey are at least as important
as measurements of their functional responses when predicting predation
on alternative prey species.

\subsection{Limitations and extensions}

\subsubsection{Can variability and mean really be separated?}

So far we equated primary prey variability to its interannual coefficient
of variation (CV). This choice needs to be discussed, especially in
the light of changes in small mammal cycles currently being observed
(e.g. \citealp{cornulier2013europe}). First, our decomposition
separates the effect of the mean primary prey density/abundance from
the effect of the CV. We chose CV because it is well-known that real
data of population abundance show a scaling between the mean and variance
of abundance. In the absence of demographic stochasticity, Taylor's
law suggests a log-variance/log-mean regression with a coefficient
of 2, which is equivalent to a CV independent of the mean. Thus, over
short enough timescales for the dynamics to be approximated by a stationary
process, and for large populations in which demographic stochasticity
has limited influence, we can assume independence of mean and CV. 

However, representing changes in variability through the CV does not
solve all problems when considering nonstationary dynamics. Recent
research on cyclic populations, for example, suggest that
mean and CV covary to some extent \citep{barraquand2014demographic,barraquand2014covariation}.
Changes in vole dynamics in e.g. the Kielder forest suggest predominantly
decreases in the mean spring vole abundance \citealp{cornulier2013europe}. 
If changes in primary prey dynamics are primarily driven by a decrease in the mean, 
cycle ``loss'' (absence of peak years) will be associated to more predation on alternative prey. 
If changes are mostly a decrease in CV, then cycle ``loss'' will usually be associated in contrast to less impact on alternative prey
(unless there are strong predator aggregative responses). 

Changes in the temporal dynamics of key herbivores might be represented in the mean-CV plane \citep{barraquand2014demographic}. This might help deciphering
the possible covariation of mean and variability over long timescales, for nonstationary dynamics, and predict the effect of changes in (cyclic) primary prey dynamics.

\subsubsection{Primary prey temporal autocorrelation}

The results we derived here, assuming mostly simple predator aggregative
responses, are valid irrespective of temporal autocorrelation in the
primary prey abundance. This follows from Birkoff's ergodic theorem,
the generalisation of the law of large numbers for autocorrelated
sequences. For example, even though rodent cycles have a famous autocorrelation
structure \citep{turchin2003cpd}, Birkoff's ergodic theorem implies
that the temporal average of a nonlinear function of rodent densities
will be equal to its expected value (average over realizations), calculated
according to the marginal probability distribution of rodent densities. For another use in ecology, 
see \citet{benaim2009persistence,barraquand2013can}. 

For the simple model with aggregative predator response, use of Birkoff's ergodic theorem is correct,
but our ability to ignore the autocorrelation critically depends on
how we relate predator numbers to primary prey densities. When delays
are added, predator numbers depend on past prey densities, and the
autocorrelation structure between primary prey densities and predator
numbers determines the average predation rate. Seasonality can generate
similar temporal patterns (\ref{ap:Seasonality}). This can be dealt with
qualitatively using simple approximations based on Taylor expansions.
Eq. \ref{eq:general_approx_total_resp} clearly shows that positive
temporal correlation between predator and prey suppresses the kill
rate on alternative prey, while negative temporal correlation increases
it. Thus, longer predator reproductive delays increase alternative
predation. But such approximations can break down when primary prey
variability is large. To a certain extent, delays might be present
in the vole-grouse-harrier example, because predators settle as a
function of current vole density (aggregative response) but have a
tendency to stay longer \citep{new2011hen,new2011modelling}. This
happens because of other alternative prey that sustain predators in
the absence of their primary prey (as pipits do for harriers), and
it can take time to deplete these other food sources. We do not think
this should alter the qualitative results in the vole-harrier-grouse
case, but are currently considering the consequences of such scenarios,
that tend to favour synchrony between prey species when predator abundance
is large (Barraquand et al. \emph{unpublished data}).

\subsubsection{Synchronization of alternative prey dynamics}

We assume throughout the paper that although breeding success and
numbers of prey 2 are affected by predation, often resulting in a
cyclic breeding success, predation is not enough to fully synchronize
prey 2 abundance with that of prey 1. This is correct for the empirical
values we possess for vole-harrier-grouse in Scotland and many other
bird-rodent systems \citep[e.g.][]{blomqvist2002indirect}, but might
not be true everywhere. In contrast, one hypothesis to explain relatively
short cycles of Fennoscandian grouse is predator sharing with cyclic
rodents \citep[the Alternative Predation Hypothesis,][]{angelstam1984role}.
Larger numbers of predators than what we considered here can lead,
in the harrier predation model, to situations where the vole and grouse
cycles synchronize (with a one-year timelag, \ref{ap:Synchronization}). Grouse
periodicity then matches vole periodicity, and this frequency entrainment
is stronger when the vole cycle exhibits strong asymmetry (as with
the MSS model). In the harrier case, densities of raptors approximately
above 1 or 2 pairs per $km^{2}$ are needed to generate such high
predation pressure, which might be difficult to obtain at large spatial
scales in most habitats. However, recent empirical work suggests that
communities of rodents are for instance synchronized by the conjugated
action of the whole predator guild \citep{korpimaki2005predator,carslake2011spatio},
which makes large predator densities feasible.

\subsubsection{More feedbacks in prey 1 - predator - prey 2 modules}

The longer snowshoe hare cycle, which is thought to be generated by
predators \citep{krebs2001drives} provides yet another nice empirical
motivation. Numerous predators such as lynx and coyotes react demographically
to snowshoe hare densities \citep{o1997numerical,o1998behavioural,o1998functional}
and inflict losses to alternative prey \citep[e.g. squirrels, sheep,][]{o1998functional,arthur2010predator}.
Hence it would be worthwhile to continue the work on asymmetric apparent
interaction modules to investigate synchronizing effects, now modeling
feedbacks between hares and predators \citep[see][]{king2001gpc},
which should allow large numbers of predators to influence the dynamics
of prey 2 when prey 1 is low. This situation can be represented prey
1$\leftrightarrow$ predator $\rightarrow$ prey 2. This would be
a logical extension to the models presented here - preliminary simulations
suggest that such delayed predator responses favour synchronization
between prey species. 

Another logical extension, to investigate apparent interactions and
synchronization between similar prey species \citep[as in][]{korpimaki2005predator,carslake2011spatio},
would be a different model structure, with also feedbacks between
prey 2 and predator dynamics (prey 1$\leftrightarrow$ predator $\leftrightarrow$
prey 2). \citet{abrams1998apparent} considered such a model with
predator-generated cycles, and they show that mutualism is possible
between prey species in the cyclic regime, which goes well with our
findings (increased mean prey abundance and decreased prey variability
are beneficial to alternative prey). However, \citet{hanski1996predation},
again with a differential equation model, inspired by interactions
between rodents and mustelids, show that apparent competition can
mediate competitive coexistence, and that a rodent species not likely
to cycle can be entrained by a competitor more susceptible to predation.
Induced cyclicity is then helping coexistence, but the presence of
prey 1 is hindering coexistence. In the case where the alternative
prey can also influence predator dynamics, our results therefore will
not always hold, so this warrants further research. 

\section{Conclusion}

Under the assumption of both asymmetric interactions (prey 1$\rightarrow$
predator $\rightarrow$ prey 2) and absence of synchronization of
prey abundances, increased mean abundance of primary prey will usually help
alternative prey species, and increased variability of primary prey
will be detrimental to alternative prey species. Unless predator aggregative response is strong, 
in which case both results can actually be reversed (Table \ref{table:results}). 
Northern rodents such as voles and lemmings, that are primary prey in many such apparent
interaction modules with ground-nesting birds, have both large maximal
abundances and large variability (CV close to 1). Their presence often
protects birds, as increases in the mean primary prey density decreases
predation on alternative prey, although in a suboptimal way, because
a less variable primary prey dynamics would usually offer better protection.
It is likely that such suboptimal protection by primary prey is felt
by other alternative prey species in a variety of ecosystems. 

\textbf{Acknowledgements}

We thank X. Lambin who encouraged us to pursue this research started in 2011. FB was funded by the Biodiversa EU program ECOCYCLES and thanks NG Yoccoz and J-A Henden for stimulating discussions. 

\newpage{}

\bibliographystyle{elsarticle-harv}
\bibliography{altpred}

\newpage{}

\appendix

\textbf{Appendices}

\section{Dealing with variation in alternative prey species}\label{ap:Dealing}

\subsection{Variation in the focal alternative prey}

So far, our 2-prey total response models are assuming prey 2 density
is roughly constant, i.e. $\mathbb{E}(T(V))$ approximates \begin{equation}
\mathbb{E}(T(V,G)|G=g)=\int_{0}^{\infty}H(v)f(v,g)\rho(v)dv\end{equation}

If $G$ is not fixed but fluctuating instead, $\mathbb{E}(T(V,G)|G)$
is a random variable. We can remark that in general, 

\begin{equation}
\mathbb{E}(T(V,G))=\int_{0}^{\infty}\int_{0}^{\infty}T(v,g)\rho_{V,G}(v,g)dvdg\end{equation}

with $\rho_{V,G}(v,g)$ the joint probability density of $V$ and
$G$. And in the simple case where they are independent (no synchrony)
we arrive at 

\begin{equation}
\mathbb{E}(T(V,G))=\int_{0}^{\infty}\int_{0}^{\infty}T(v,g)\rho_{V}(v)\rho_{G}(g)dvdg=\int_{0}^{\infty}\mathbb{E}(T(V,G)|G=g)\rho_{G}(g)dg\end{equation}

So we just need to integrate once more the average total response,
this time with respect to grouse variation. This shows \emph{why}
our simple approximation is working, despite the variation in grouse
density in the numerical models. 

We might even take into account synchrony between grouse and vole
using some functional form $G=G_{0}+\beta V$. That is a deterministic
link expressing an extraordinarily strong level of synchrony. A more
realistic model could assume that grouse is linearly related to vole
density \emph{on average}, e.g. $G|V\sim N(G_{0}+\beta V,\sigma_{G|V}^{2})$,
assuming here a normal probability distribution conditional to vole
density (we recover independence between $V$ and $G$ when $\beta=0$).
This would result in the conditional p.d.f. $\rho_{G|V=v}(g;v)$ being
known, and then we can write

\begin{equation}
\mathbb{E}(T(V,G))=\int_{0}^{\infty}\int_{0}^{\infty}H(v)f(v,g)\rho_{G|V=v}(g;v)\rho_{V}(v)dvdg\end{equation}

where all the elements are known, or again,

\begin{eqnarray}
\mathbb{E}(T(V,G))=\int_{0}^{\infty}H(v)\left(\int_{0}^{\infty}f(v,g)\rho_{G|V=v}(g;v)dg\right)\rho_{V}(v)dv \\
= \int_{0}^{\infty}H(v)\mathbb{E}(f(V,G)|V=v)\rho_{V}(v)dv\end{eqnarray}

where the quantity between parentheses is the expectation of the functional
response on the grouse as a function of grouse density, conditional
on vole density. That expectation is possible to compute, at least
by numerical integration. Note: Similar decompositions using conditional
expectations could help dealing with parameter uncertainty. A difficulty
not addressed here is that synchrony between $V$ is $G$ usually
working with a lag (\ref{ap:Synchronization}), thus the arguments above might need
to apply to some function of $G$.

\subsection{Variation in another prey species}

The above arguments can be extended for variation in other prey species
than the primary prey or the alternative prey of interest. Thinking
of the harrier example, let us consider a density of pipits $P_{t}$,
that varies over time, independently of $V_{t}$. Assuming $G_{t}$
is constant (e.g. applying the arguments of the previous subsection),
the total response writes

\begin{equation}
\mathbb{E}(T(V,P))=\int_{0}^{\infty}\int_{0}^{\infty}T(v,p)\rho_{V,P}(v,p)dvdp\end{equation}

Due to independence, we arrive at

\begin{equation}
\mathbb{E}(T(V,P))=\int_{0}^{\infty}\int_{0}^{\infty}T(v,p)\rho_{V}(v)dv\times\rho_{P}(p)dp\end{equation}

which is then equal to

\begin{equation}
\mathbb{E}(T(V,P))=\int_{0}^{\infty}\mathbb{E}(T(V,P|P=p))\rho_{P}(p)dp\end{equation}

We therefore show that considering pipit density, provided marginal
probability distributions of prey densities can be considered independent,
only requires one additional integration over the whole pipit probability
distribution. 

\section{Inclusion of optimal foraging behaviour in the functional response}\label{ap:foraging}

Here we modified a classic type-II multispecies functional response
model to allow for optimal foraging - this is generally thought to
stabilize such models and enlarge the spectrum for prey species coexistence
\citep{Krivan1996265,krivan1999optimal}. We follow closely \citet[chap. 2]{fryxell1998individual}
and the initial type II response, similar to eq. (5) in the main text,
comes from \citet{hanski1996predation}, hereafter referred to as HH96.

The basic model of foraging theory, i.e. the diet choice model \citep{stephens1986ft},
predicts that prey items should be ranked according to their relative
profitabilities $e_{i}/h_{i}$ (energetic content/handling time) and
the least profitable prey is ignored until the most profitable type
drops below some critical density threshold. For convenience here
we assume the most profitable prey type is 1, i.e. $e_{1}/h_{1}>e_{2}/h_{2}$.

In comparison, the type II functional response (sometimes called {}``shared
predation'' response, \citealp{norrdahl2000predators}) assumes a heavier predation on the less profitable
prey when the main prey is abundant. Let us assume that $\gamma_{i}$
is the probability of accepting the prey item $i$ upon encounter
and modify the functional response. Using an encounter (discovery)
rate parameter $a$, the functional response (predator intake rate)
now reads classically

\[
\frac{a\gamma_{i}V_{i}}{1+a(\gamma_{1}h_{1}V_{1}+\gamma_{2}h_{2}V_{2})}\]

where $V_{1}$ and $V_{2}$ are the densities of the prey one and
two, respectively. We have $\gamma_{1}=1$ and $\gamma_{2}=1$ if
$V_{1}>\eta_{1}$, $0$ if $V_{1}<\eta_{1}$. In the classic theory
\citet{stephens1986ft,fryxell1998individual,Krivan1996265}, the long-term
individual intake is written

\[
w=\frac{a(\gamma_{1}e_{1}V_{1}+\gamma_{2}e_{2}V_{2})}{1+a(\gamma_{1}h_{1}V_{1}+\gamma_{2}h_{2}V_{2})}\]

Through the derivation of $w$ w.r.t. $\gamma_{1}$ and $\gamma_{2}$,
we find the critical main prey density threshold is $\eta_{1}=\frac{e_{2}}{a(e_{1}h_{2}-e_{2}h_{1})}$
\citet{stephens1986ft,fryxell1998individual,Krivan1996265}. This
step-like formulation is not practical and is not likely to be observed
in nature (partial preferences, \citealp{berec2000mmp}) so that \citet[chap. 2]{fryxell1998individual}
propose a sigmoid function instead, $\gamma_{2}=\frac{\exp(z\eta_{1})}{\exp(zV_{1})+\exp(z\eta_{1})}$
($z$ is a smoothing parameter controlling the steepness of the sigmoid).
We therefore included optimal diet choice, at the expense of two additional
parameters only
\begin{enumerate}
\item The ratio $k$ of energetic content of two prey items
\item The steepness $z$ of the partial preference curve ($\gamma_{2}=f(V_{1})$)
\end{enumerate}
The results in the figure show that the inclusion of optimal foraging
tend to produce exponential-like functional response on prey 2 as
a function of prey 1 density (bottom right panel). 

\setcounter{figure}{0}

\begin{center}
\begin{figure}
\begin{centering}
\includegraphics[width=7cm]{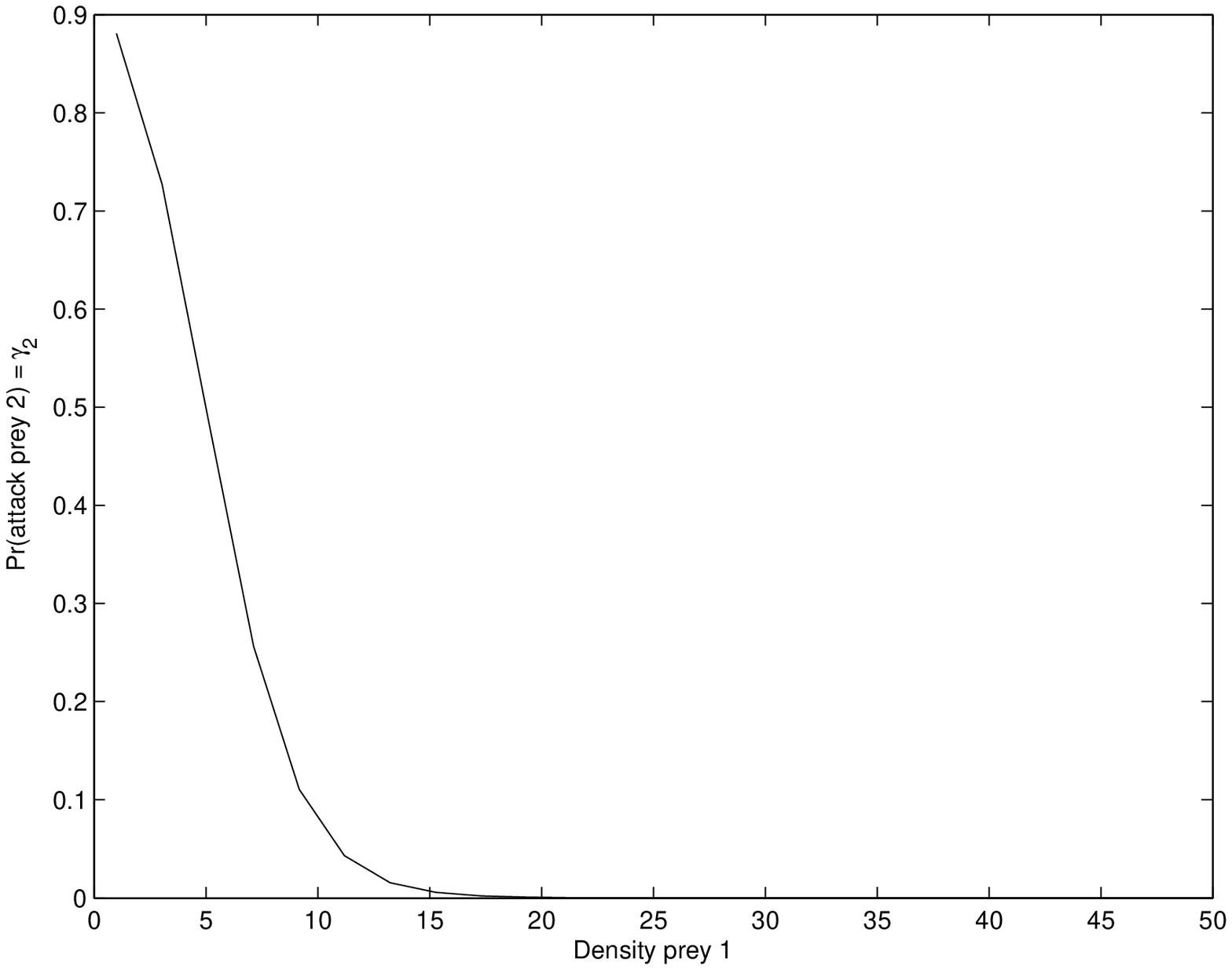}
\includegraphics[width=8cm]{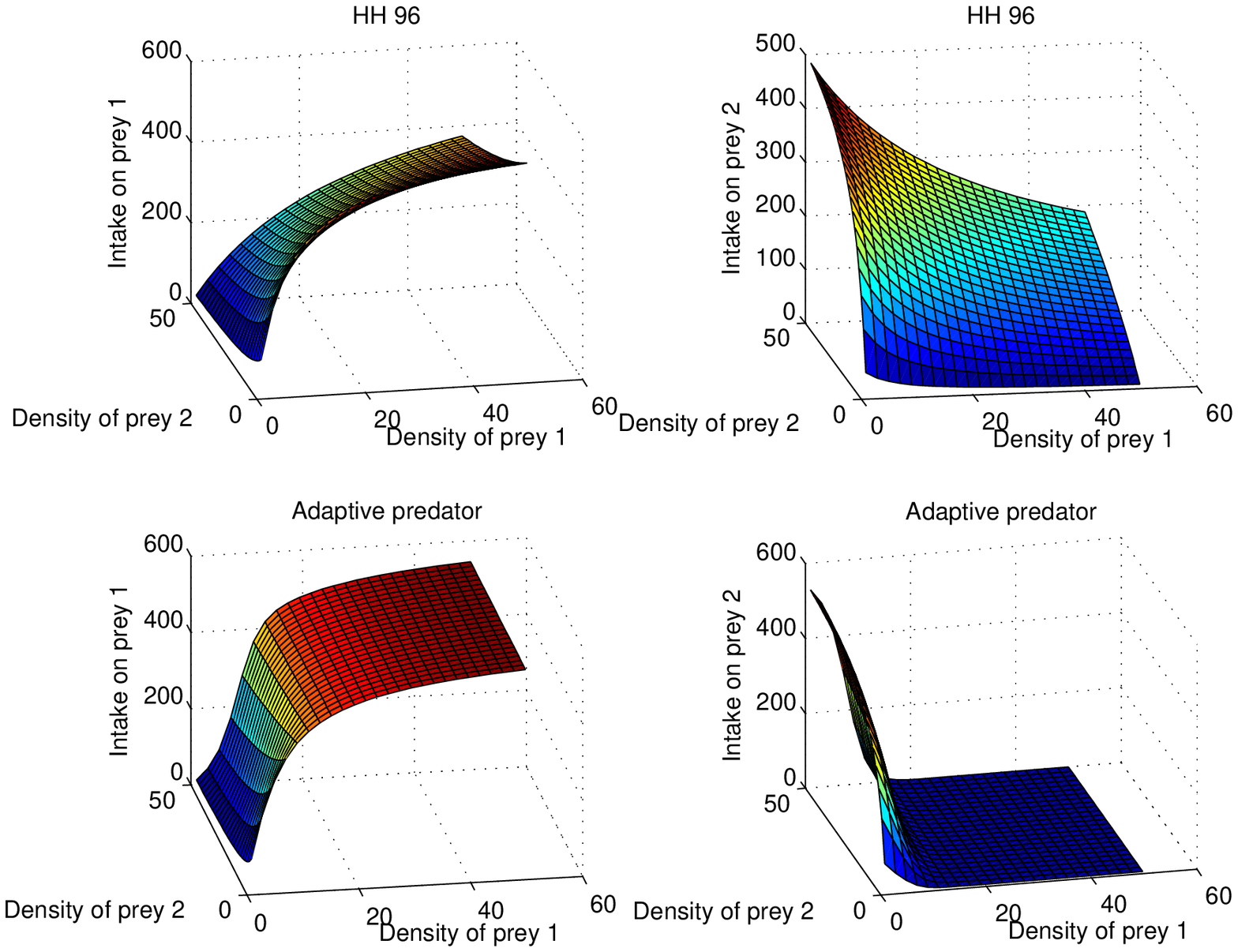}
\par\end{centering}

\caption{Left: Illustration of how the probability of attack of alternative
prey should vary with main prey density, for an optimal predator preying
mainly on rodents (prey 1) and alternatively on gamebird (prey 2).
There is a slight threshold (inflection point) but it occurs at quite
low rodent densities. Resulting functional response on the right (adaptive
predator) and comparison to a type II functional response (HH96, from
\citet{hanski1996predation}'s apparent competition model). Parameters: $D_1 = 5.0, D_2 = 10.0, h=1/600, a=\frac{1}{Dh}, k=e_1/e_2=2, z=0.5$}

\end{figure}

\par\end{center}

\section{Correlations between average population size and average breeding
success / natural mortality in alternative prey}\label{ap:Correlations}

Let us assume the alternative prey has population dynamics of the
form $G_{t+1}=G_{t}R(V_{t})/k(G_{t})$, with a maximum per capita
growth rate $R(V_{t})=B(V_{t})+D(V_{t})$ (either maximum breeding
success or natural mortality can be affected by primary prey density)
and the form of regulation $k(G_{t})=e^{\lambda G_{t}}$ (Ricker),
or $(1+G_{t}/K)^{\beta}$ (Hassell), or $(1+(G_{t}/K)^{\gamma})$
(Maynard-Smith). By definition, in a regulated population the long-term
growth rate has to be zero. The only assumption made here is that primary prey density affects mostly the maximal per capita growth
rate, and not the density-dependent structure (this is only an approximation,
but makes sense if one thinks density-dependence is mostly set by
resources not predators). 

Because $\ln(G_{t+1}/G_{t})=\ln(R(V_{t}))-\ln(k(G_{t})),$ we arrive
at the equation $\mathbb{E}\ln(R(V))=\mathbb{E}\ln(k(G))$ \citep[some uses of this technique in][]{barraquand2013can}. 
In the case of the Ricker model, this even transforms into a more direct
relationship between population size and the expected log-max pop
growth rate $\mathbb{E}(G)=\frac{\mathbb{E}\ln(R(V))}{\lambda}$.
If either $B(V_{t})$ or $D(V_{t})$ is close to 0.5 - let us assume
$D(V_{t})$ is, as in the grouse case - and $B(V_{t})$ has small
fluctuations, we simplify $\ln(1+B(V_{t})-0.5)\approx B(V_{t})-0.5$,
and thus approximately $\mathbb{E}(G)\propto\mathbb{E}(B(V)+\textrm{constant})$.
If we also have $\mathbb{E}(B(V))\propto\mathbb{E}(T(V))$ (proportionality
of the breeding success to the expected total response), which should
be the case whenever $G$ does not fluctuate too much, we then obtain
$\mathbb{E}(G)\propto\mathbb{E}(T(V))$. This is what we found with
the vole-harrier-grouse model. This relationship works in a number
of models for $k(G_{t})$ (e.g. Ricker, Hassell, Maynard-Smith).

\section{Sensitivity analysis of the average total response to mean and coefficient
of variation in primary prey density}\label{ap:Sensitivity}

\paragraph{Gamma distribution reminder}

The probability density function is $\rho(x)=\frac{x^{k-1}e^{-x/\theta}}{\theta^{k}\Gamma(k)}=\beta^{k}\frac{x^{k-1}e^{-\beta x}}{\Gamma(k)}$
with $(k,\beta)=(k,1/\theta)$ and $\mathbb{E}(X)=k\theta,\mathbb{V}(X)=k\theta^{2}$.
Therefore, the coefficient of variation is $c=1/\sqrt{k}$, and the
shape parameter is $k=1/c^{2}$.

\subsection{Computation of the general integral for the total response}

The integration uses the fact that $\int_{0}^{\infty}\rho(x)dx=1$
because $\rho$ is a pdf on $[0,+\infty)$ to simplify the integral.
Let's call the integral $\bar{T}=\int_{0}^{\infty}\rho(x)T(x)dx$.
We have

\begin{equation}
\bar{T}=\int_{0}^{\infty}\rho(x)T(x)dx=\int_{0}^{\infty}\beta^{k}\frac{x^{k-1}e^{-\beta x}}{\Gamma(k)}Ax^{l}e^{-\lambda x}dx\end{equation}

\begin{equation}
\bar{T}=A\frac{\beta^{k}}{\Gamma(k)}\int_{0}^{\infty}x^{l+k-1}e^{-(\lambda+\beta)x}dx\end{equation}

\begin{equation}
\bar{T}=A\frac{\beta^{k}\Gamma(k+l)}{\Gamma(k)(\lambda+\beta)^{k+l}}\underbrace{\int_{0}^{\infty}\frac{(\lambda+\beta)^{k+l}}{\Gamma(k+l)}x^{l+k-1}e^{-(\lambda+\beta)x}dx}_{\int_{0}^{\infty}\rho_{k+l;\lambda+\beta}(x)dx=1}\end{equation}

We therefore have in the general case \begin{eqnarray}
\bar{T} & = & A\frac{\beta^{k}\Gamma(k+l)}{\Gamma(k)(\lambda+\beta)^{k+l}}\label{eq:general_av_total_resp}\end{eqnarray}

Let's consider a simple case

\paragraph{Case $l=1$, $H(V)=\alpha V$: linear aggregative response}

Because $\Gamma(k+1)=k\Gamma(k)$, we obtain \begin{eqnarray}
\bar{T} & = & A\frac{k}{\beta}\left(\frac{\beta}{\lambda+\beta}\right)^{k+1}\end{eqnarray}

With the other formulation for the gamma distribution, denoting $\lambda=1/\eta$
($\eta$ is the vole density at which the individual predation rate
on the alternative prey drops to $e^{-1}=37\%$ of its initial value),
we obtain

\begin{equation}
\bar{T}=Ak\theta\left(\frac{\eta}{\eta+\theta}\right){}^{k+1}=\varphi_{1}(k,\theta)\end{equation}

NB: $\frac{\beta}{\lambda+\beta}=\frac{\eta}{\eta+\theta}$ with $\beta=1/\theta$. 

Or, writing the mean rodent density $m=k\theta$ and $c=1/\sqrt{k}$
the coefficient of variation of rodent density (the relevant variable
for rodent variability)

\begin{equation}
\bar{T}=Am\left(\frac{\eta}{\eta+mc^{2}}\right){}^{1/c^{2}+1}=\varphi_{2}(k,c)\end{equation}

Or again, since $k=1/c^{2}$ , a more practical formulation for computations
is

\begin{equation}
\bar{T}=Am\left(\frac{\eta}{\eta+m/k}\right){}^{k+1}=\varphi_{3}(k,m)\end{equation}

We use only the last version, in the following calculations so we
drop the subscript 3.

\paragraph{Case $l=0$: constant predator density}

We then have $H(V)=H_{0}$ and $A=H_{0}f_{0}$ in the integral. 

\begin{eqnarray}
\bar{T} & = & A\frac{\beta^{k}\Gamma(k+l)}{\Gamma(k)(\lambda+\beta)^{k+l}}=A\frac{\beta^{k}}{(\lambda+\beta)^{k}}=A\left(\frac{\eta}{\eta+\theta}\right)^{k}\end{eqnarray}

Thus, 

\begin{equation}
\bar{T}=A\left(\frac{\eta}{\eta+m/k}\right)^{k}=\psi(k,m)\end{equation}

Of course, we can have any combination of the two previous cases,
such as $H(V)=H_{0}+\alpha V$ ; due to the additivity of the integral
this leads to

\begin{eqnarray}
\bar{T} & = & H_{0}f_{0}\left(\frac{\eta}{\eta+m/k}\right)^{k}+\alpha f_{0}m\left(\frac{\eta}{\eta+m/k}\right){}^{k+1}\nonumber \\
 &  & =f_{0}\left(\frac{\eta}{\eta+m/k}\right)^{k}\left(H_{0}+\frac{\eta\alpha m}{\eta+m/k}\right)\end{eqnarray}

Below, both cases are interpreted separately, while B11 will be evaluated
at a later date.

\subsection{Sensitivity analysis of the averaged total response $\bar{T}$}

We computed partial derivatives with respect to the average rodent
density $m$ and its coefficient of variation $c$ (through parameter
$k=1/c^{2}$).

\subsection{Effect of the rodent mean}

\subsubsection{Case $l=1$ ($\approx$ with aggregative response)}

We need to compute $\frac{\partial\varphi}{\partial m}$. Let's write
$A=\alpha f_{0}$. We then have \begin{equation}
\varphi(k,m)=Am\left(\frac{\eta}{\eta+m/k}\right)^{k+1}=Am\frac{\partial}{\partial k}e^{(k+1)\ln(\frac{\eta}{\eta+m/k})}=Ame^{u(m)}\label{eq:phi8-1}\end{equation}

which leads to

\begin{equation}
\frac{\partial\varphi}{\partial m}=Am\frac{\partial}{\partial m}(e^{u(m)})+Ae^{u(m)}=Ae^{u(m)}(mu'(m)+1)\end{equation}

with $u(m)=(k+1)[\ln(\eta)-\ln(\eta+m/k)]$. Thus,

\begin{equation}
u'(m)=(k+1)(-1)\frac{\frac{1}{k}}{\eta+m/k}=\frac{-k-1}{k\eta+m}\end{equation}

Finally, 

\begin{equation}
mu'(m)+1=\frac{-km-m+k\eta+m}{k\eta+m}=\frac{k(\eta-m)}{k\eta+m}\end{equation}

Therefore, \begin{equation}
\frac{\partial\varphi}{\partial m}>0\iff m<\eta\end{equation}

The mean rodent density $m$ has a positive effect on the averaged
total response $\bar{T}=\varphi(m,k)$ when $m$ is small, and a negative
effect when $m$ is large (maximum effect at $m=\eta$ for the averaged
total response, $\eta$ being also the maximum the non-averaged total
response $T(V)$ here).

\subsubsection{Case $l=0$ ($\approx$ without aggregative response)}

We need to compute $\frac{\partial\psi}{\partial m}$. Let's write
$A=H_{0}f_{0}$. We have 

\begin{equation}
\psi(k,m)=A\left(\frac{\eta}{\eta+m/k}\right)^{k}=Ae^{k\ln(\eta)}e^{-k\ln(\eta+m/k)}=A'e^{u(m)}\end{equation}

with $A'=Ae^{k\ln(\lambda)}$ and $u(m)=-k\ln(\lambda+m/k)$ which
leads to $u'(m)=-k\frac{1/k}{\eta+m/k}=\frac{-1}{\eta+m/k}$. Because
$\frac{\partial e^{u(m)}}{\partial m}=u'(m)e^{u(m)}$, we obtain $\frac{\partial\psi}{\partial m}<0$.
So the effect of increasing mean rodent density is to decrease the
average kill rate when $l=0$ (no aggregative response), and corresponds
to the monotonically decreasing total response in that case.

\subsection{Effect of variability}

Now we have to look at how the average kill rate is influenced by
the coefficient of variation $c$. The effect of more variability
can be positive on the kill rate, but at large mean $m$ only where
there is an aggregative response. Note here that $\frac{\partial\varphi}{\partial c}=\frac{\partial\varphi}{\partial k}\frac{\partial k(c)}{\partial c}$,
where $k=1/c^{2}$, and hence $\frac{\partial k}{\partial c}<0$.

\subsubsection{Case $l=1$ ($\approx$ with aggregative response)}

\begin{eqnarray}
\varphi(k,m) & = & Am\left(\frac{\eta}{\eta+m/k}\right){}^{k+1}=Am\left(\frac{1}{1+x(k)}\right){}^{k+1}\nonumber \\
 &  & =Am\exp(-(k+1)\ln(1+x(k)))\end{eqnarray}

with $x(k)=\frac{m}{k\eta}$. Then we obtain $\frac{\partial\varphi}{\partial k}=Am\frac{\partial}{\partial k}e^{-(k+1)\ln(1+x(k))}$.
Let's denote $u(k)=-(k+1)\ln(1+x(k))$; we have $\frac{\partial}{\partial k}e^{u(k)}=u'(k)e^{u(k)}$.
So the core of the problem is to find 

\begin{equation}
u'(k)=\frac{\partial(-k-1)}{\partial k}\ln(1+x(k))-(k+1)\left(\frac{x'(k)}{1+x(k)}\right)\end{equation}

\begin{equation}
u'(k)=-\ln(1+x(k))-(k+1)\left(\frac{-\frac{m}{k^{2}\eta}}{1+x(k)}\right)\end{equation}

\begin{equation}
u'(k)=-\ln(1+\frac{m}{k\eta})+\frac{\frac{m}{k}(k+1)}{k\eta+m}\end{equation}

Let's assume that $\frac{m}{k\eta}<<1$, which seems likely ($m<\eta$
implies mean rodent density is lower than the threshold at the functional
response drops to $e^{-1}=37\%$ of its maximum, and $k>1$). In that
case, we have $\ln(1+\frac{m}{k\eta})\approx\frac{m}{k\eta}$ which
then leads to 

\begin{eqnarray}
u'(k) & = & -\frac{m}{k\eta}+\frac{\frac{m}{k}(k+1)}{k\eta+m}=\frac{-m(k\eta+m)+\frac{m}{k}k\eta(k+1)}{k\eta(k\eta+m)}\nonumber \\
 &  & =\frac{-mk\eta-m^{2}+mk\eta+m\eta)}{k\eta(k\eta+m)}=\frac{m(-m+\eta)}{k\eta(k\eta+m)}\end{eqnarray}

So $u'(k)<0$ iff $m>\eta$ (assuming the approximation works). We
know $\frac{\partial\varphi}{\partial c}=\left(Amu'(k)e^{u(k)}\right)\frac{\partial k}{\partial c}$
and $\frac{\partial k}{\partial c}=\frac{-2}{c^{3}}<0$ so 

\begin{equation}
\frac{\partial\varphi}{\partial c}>0\Leftrightarrow m>\eta\end{equation}

Or, in other words, there is a positive effect of rodent variability
on the kill rate of grouse at large average rodent density.
The results presented here are actually similar to those that can
be obtained using ``small-noise'' approximations (i.e. with Taylor
expansions of nonlinear functions). 

Now the question that remain is: can there be other `negative' effects
of variability at not only very high, but also very low rodent density?
This may occur when the aggregative response is accelerating, i.e.
$H(V)=\alpha_{l}V^{l}\Rightarrow T(V)=A_{l}V^{l}\exp(-\lambda V)$.
To investigate this question, we need to redo these computations with
a power $l>1$ in the integral, and this is done in \ref{ap:lsup1}.

\subsubsection{Case $l=0$ ($\approx$ without aggregative response)}

We have $\psi(k,m)=H_{0}f_{0}e^{k\left[\ln(\eta)-\ln(\eta+m/k)\right]}$.
Let us denote $u(k)=k\left[\ln(\eta)-\ln(\eta+m/k)\right]$. Then

\begin{equation}
\frac{\partial u}{\partial k}=\ln(\eta)-\ln(\eta+m/k)-\frac{k\frac{1}{k}}{\eta+m/k}=\ln(\frac{\eta}{\eta+m/k})-\frac{1}{\eta+m/k}=-\ln(1+x)-\frac{1}{\eta}\frac{1}{1+x}\end{equation}

with $x=\frac{m}{k\eta}$. $x>0$ so $\ln(1+x)>0$ which means $u'(k)<0$
and $\frac{\partial\psi}{\partial k}=\frac{\partial}{\partial k}e^{u(k)}=u'(k)e^{u(k)}<0$.
Consequently, $\frac{\partial\psi}{\partial c}>0$, and increase in
rodent variability, for a constant mean, will increase the kill rate
on the grouse. We recover the negative effects on grouse of rodent
variability that can be found using simply Jensen's inequality, in
the case of constant predator numbers.

\subsection{Effect of variability, case $l>1$, accelerating total response at
low rodent densities\label{ap:lsup1}}

Here we show that positive effects of vole variability on grouse kill
rates are unlikely at very low mean rodent density, despite the convex
shape of the total response curve ($l>1$) suggesting so. We rewrite
the general eq. \ref{eq:general_av_total_resp} for the averaged total
response when assuming the flexible $v^{l}e^{-\lambda v}$ form:

\[
\bar{T}=\alpha f_{0}\frac{\beta^{k}\Gamma(k+l)}{\Gamma(k)(\lambda+\beta)^{k+l}}\]

we have the relations $\beta=\frac{1}{\theta}=\frac{k}{m},\lambda=\frac{1}{\eta}$.
Hence, using also $\Pi(k,l)=\frac{\Gamma(k+l)}{\Gamma(k)}=k(k+1)...(k+l-1)$,
we arrive at

\begin{equation}
\bar{T}=\alpha f_{0}\frac{(k/m)^{k}\Pi(k,l)}{(\lambda+k/m)^{k+l}}=\varphi_{4}(k,m)\end{equation}

Taking the derivative with respect to $k$ here is not easy. Because
we are concerned with very low $m$ values, we will make the approximation
that $k/m>>\lambda$. This is equivalent to the approximation $\frac{m}{k\eta}<<1$
done for the case $l=1$. There we arrive at 

\[
\bar{T}=A(k/m)^{-l}\Pi(k,l)=\phi(k,m)\]
There, we have \begin{equation}
\frac{\partial\phi}{\partial k}=A\left[\Pi(k,l)\frac{\partial}{\partial k}\left((k/m)^{-l}\right)+(k/m)^{-l}\frac{\partial\Pi(k,l)}{\partial k}\right]\end{equation}

Let's compute the logarithmic derivative of $\Pi(k,l)$, 

\begin{equation}
\ln(\Pi(k,l))=\ln(k)+\ln(k+1)+...+\ln(k+l-1)\end{equation}

\begin{equation}
\frac{\partial\ln(\Pi(k,l))}{\partial k}=\underbrace{\frac{1}{k}+\frac{1}{k}+...+\frac{1}{k}}_{k+l\; times}=(k+l)\frac{1}{k}\end{equation}

From the chain rule, \begin{equation}
\frac{\partial\ln(\Pi(k,l))}{\partial k}=\frac{\partial\Pi(k,l)}{\partial k}\frac{1}{\Pi(k,l)}\end{equation}

which implies

\begin{equation}
\frac{\partial\Pi(k,l)}{\partial k}=\Pi(k,l)\frac{\partial\ln(\Pi(k,l))}{\partial k}=\Pi(k,l)\frac{k+l}{k}\end{equation}

Another useful relationship is \begin{equation}
\frac{\partial}{\partial k}\left(\left(\frac{k}{m}\right){}^{-l}\right)=\frac{-l}{m}\left(\frac{k}{m}\right)^{-l-1}=\frac{-l}{m}\left(\frac{m}{k}\right)^{l+1}\end{equation}

Therefore, 

\begin{equation}
\frac{\partial\phi}{\partial k}=A\Pi(k,l)\left[\frac{-l}{m}\left(\frac{m}{k}\right)^{l+1}+\left(\frac{m}{k}\right)^{l}\frac{k+l}{k}\right]\end{equation}

\begin{equation}
\frac{\partial\phi}{\partial k}=A\Pi(k,l)\left(\frac{m}{k}\right)^{l}\left[\frac{-l}{m}\frac{m}{k}+\frac{k+l}{k}\right]=A\Pi(k,l)\left(\frac{m}{k}\right)^{l}>0\label{eq:approx_doubtful}\end{equation}

Thus for low $m$, $\frac{\partial\phi}{\partial k}>0\iff\frac{\partial\phi}{\partial c}<0$,
increased prey variability decreases the total response on the alternative
prey, which is {}``good'' from the alternative prey point of view.
Confirmed with numerical simulations (Fig. 4 in the main text).

\subsection{Negative correlation between predator and prey}

Let us consider here $H(V)=H_{0}+\alpha V$, with $\alpha<0$, which
is an approximate way of representing a (strong) negative temporal
correlation. We then obtain $\frac{\partial\bar{T}}{\partial c}=\frac{\partial\psi}{\partial c}+\frac{\partial\varphi}{\partial c}$.
We know $\frac{\partial\psi}{\partial c}>0$ from section B.5. Previously,
with $\alpha>0$, we found $\frac{\partial\varphi}{\partial c}>0\Leftrightarrow m>\eta$.
Thus now $\frac{\partial\varphi}{\partial c}>0\Leftrightarrow m<\eta$.
Hence when $m<\eta$, it is clear that $\frac{\partial\bar{T}}{\partial c}>0$.
The averaged total response increases with primary prey variability. 

When $m>\eta$ however (large mean primary prey density, which is
improbable), it seems possible that $\frac{\partial\bar{T}}{\partial c}<0$
whenever $-\frac{\partial\varphi}{\partial c}>\frac{\partial\psi}{\partial c}$.

\section{Seasonality in the cycle, or how to introduce delays in the aggregative
response}\label{ap:Seasonality}

A case where the aggregative response assumption (predator numbers
are a function of the current primary density) is not possible is
obviously when seasonality implies delays. 
In the case of rodent/raptor/ground-nesting bird systems, the (cyclic)
autocorrelation structure in rodent dynamics can then matter. 
We did not consider the fact that predators settle in early spring \citep[as in][for lemming predators]{gilg2003cyclic};
when rodents decrease rather than increase over the summer, one can sometimes
observe heavy predation pressure on alternative prey at the end of
the summer. So far, all density values had to be thought of as end-of-spring/early-summer
values. Now let us consider a two-season model with censuses for a
rodent (e.g. vole)
\begin{enumerate}
\item At the end of winter $V_{i,1}$
\item At the end of summer $V_{i,2}$
\end{enumerate}
The functional response is not affected by any delay, i.e. $f(V_{i,s})=f_{0}e^{-\lambda V_{i,s}}$.
In contrast, the aggregative response is assumed to depend on spring
density only: 
\begin{enumerate}
\item $H(V_{i,1})=\alpha V_{i,1}$
\item $H(V_{i,2})=\alpha V_{i,1}$
\end{enumerate}
This is simplest case of aggregative response (more complicated functions
would not affect the main conclusions). In that case, we have

\begin{equation}
\bar{T}=\frac{1}{n}\sum_{i=1}^{n}\frac{1}{2}\left(T(V_{i,1})+T(V_{i,2})\right)=\frac{1}{n}\sum_{i=1}^{n}\frac{1}{2}\alpha V_{i,1}\left(f(V_{i,1})+f(V_{i,2})\right)\end{equation}

We have then two different cases to examine with respect to the seasonality in vole
dynamics.

\subsection{Rodent crashes occur mostly in winter, but declines are nonetheless possible in summer} 

We model the dynamics of $V_{i,1}=V_{i}$ (winter-driven population cycles) and write $V_{i,2}=R_{S}V_{i,1}$
with $R_{S}$ the summer growth rate. We obtain the same model as
before except

\begin{equation}
\bar{T}=\frac{1}{n}\sum_{i=1}^{n}\alpha V_{i}\bar{f}(V_{i})\end{equation}

with a corrected functional response on the grouse that can typically
be larger due to the decline in vole availability during the summer:

\begin{equation}
\bar{f}(V)=\frac{f_{0}}{2}\left(e^{-\lambda V}+e^{-\lambda R_{S}V}\right)\end{equation}

If $R_{S}$ is small (e.g. 1/4), for most $V$ values that are not
too large, we typically have $e^{-\lambda R_{S}V}\approx1>>e^{-\lambda V}$
(simply because the negative exponential function is decreasing, and
relatively steep at the beginning). Thus, seasonality can have an
important impact and bring the alternative prey population to levels
of predation that would be unexpected given the yearly average of
the vole density. However, the following case generates even stronger predation.

\subsection{Rodent crashes always occur during summer: delays appear and strongly increase
the level of predation}

Let us consider now that rodent population crashes occur during the course of the summer (as in collared lemmings,
\citealt{gilg2003cyclic}), which means early summer densities can reach temporarily high levels, but if end-of-winter densities are large, then end-of-summer densities
are low. Winter density is then treated as a fixed factor (or random
growth rate), $R_{W}$ multiplied by the summer density. 
$V_{i,2}=V_{i}$ becomes the dynamic variable, and we have $V_{i,1}=R_{W}V_{i-1,2}=R_{W}V_{i-1}$. 
Then we obtain

\begin{eqnarray}
\bar{T} & = & \frac{1}{n}\sum_{i=2}^{n+1}\frac{1}{2}\alpha V_{i,1}\left(f(V_{i,1})+f(V_{i,2})\right)=\frac{1}{n}\sum_{i=2}^{n+1}\frac{1}{2}\alpha R_{W}V_{i-1,2}\left(f(R_{W}V_{i-1,2})+f(V_{i,2})\right)\nonumber \\
 & = & \frac{1}{n}\sum_{i=2}^{n+1}\frac{1}{2}\alpha R_{W}V_{i-1}\bar{f}(V_{i},V_{i-1})\end{eqnarray}

with $\bar{f}(V_{i},V_{i}-1)=f(R_{W}V_{i-1})+f(V_{i})$. 

This model is interesting only in the case of wide amplitude fluctuations,
with peaks followed by a severe crash. In this case, we can have a
very high beforecrash autumn density $V_{i-1}$, generating very high
abundance in the next spring, $R_{W}V_{i-1}$. Let's assume for the
sake of simplicity $R_{W}=1$ (or larger). 

Following that logic, $f(V_{i-1})$ at the end of winter/spring beginning
is negligible (because $f$ is decreasing). But then, at the end of
summer the predation rate is $f(V_{i})$, which can be quite high
($V_{i}<<V_{i-1}$ if the crash happens in between) and this is multiplied
by the number of predators proportional to $V_{i-1}$, which is still
very high. 

Therefore, a delay in the aggregative response will generate much
more predation when crashes occur in summer than in winter. To see
how much, we have to use a model where we can manipulate amplitude
easily and have control of the periodicity. 

A possible candidate is the sequence $V=(v_{1},v_{2},v_{3},v_{1},v_{2},v_{3},v_{1},v_{2},v_{3},...)$
where $v_{1}=v,v_{2}=Rv,v_{3}=R^{2}v$ (geometric growth with annual
growth rate $R$ combined with periodic crashes). This would lead for the average
kill rate \begin{equation}
\bar{T}=\frac{\alpha}{2n}\sum_{i=2}^{n+1}\frac{1}{3}\left(v_{1}f(v_{2})+v_{2}f(v_{3})+v_{3}f(v_{1})\right)=\frac{\alpha}{6}\left(vf(Rv)+Rvf(R^{2}v)+R^{2}vf(v)\right)\end{equation}

instead of the winter-crash case described in the previous section

\begin{equation}
\bar{T}=\frac{\alpha}{2n}\sum_{i=2}^{n+1}\frac{1}{3}\left(v_{1}f(v_{1})+v_{2}f(v_{2})+v_{3}f(v_{3})\right)=\frac{\alpha}{6}\left(vf(v)+Rvf(Rv)+R^{2}vf(R^{2}v)\right)\end{equation}

We can just compare the two averages $vf(Rv)+Rvf(R^{2}v)+R^{2}vf(v)$
and $vf(v)+Rvf(Rv)+R^{2}vf(R^{2}v)$
\begin{itemize}
\item Summer crash case: the average is roughly $T_{SC}=vf(Rv)+Rvf(R^{2}v)+R^{2}vf(v)\approx R^{2}vf(v)$
(other values are negligible provided $R$ is quite large, and most
predation occurs in the year with the crash)
\item Winter crash case: $T_{WC}=vf(v)+Rvf(Rv)+R^{2}vf(R^{2}v)\approx Rvf(Rv)$
(other values are negligible because $v$ is close to zero and $f(R^{2}v)$
as well, most predation occurs in the intermediate year: remember
the total response in that case has a maximum for intermediate vole
values)
\item We know $f(Rv)<f(v)$ because $R>1$ and $f$ is decreasing. 
We can even assume $f(Rv)<<f(v)$ if $R>>1$. We therefore obtain $T_{WC}\approx Rvf(Rv)<<Rvf(v)$.
Then, $T_{WC}\approx Rvf(Rv)<<R^{2}vf(v)\approx T_{WS}$ (since $R$
is in general $>10$ for these species of rodents or very variable
key herbivore). We can generalize the result to cycles longer than
3 years.
\end{itemize}

Conclusion: rodent dynamics with a summer crash is worse for alternative
prey than rodent dynamics with a winter crash. Crash being understood
here as a sharp reduction in abundance, visible in the annual time
series (either spring-to-spring or autumn-to-autumn). 

Considering rodent dynamics with a winter crash (the classic case
when thinking about voles, but not lemmings), maximal predation will likely happen
during the intermediate year: there are not enough predators at low
vole density to do any damage, and the numerous predators in the peak
year have very low predation on alternative prey because they prey
mainly on abundant rodents. However, when instead there is a summer crash,
most predation occurs right after the crash; when there are
both numerous predators and no rodents, which means there is extremely
heavy alternative predation.

This suggests that nest predation might be worse for ground-nesting
birds when the main rodent is a lemming rather than a vole, because
lemmings often crash during the summer while they increase in winter.
Note: all this is conditional to predators having a strong aggregative
response, although this is usually the case when it comes to lemmings
in the high North \citep{gilg2003cyclic,gilg2006functional}. 

Similar effects will occur when we allow for a predator numerical
response, but on a multiannual rather than seasonal scale. The basic
mechanism is the same: a negative temporal correlation between primary prey and predator densities. 
Most of the predation will then occur in the crash
phase, because the predators stay quite numerous and are without anything
else to eat. Unfortunately, because predation can be very strong
in this case, this implies that the simplifying assumption that vole
and grouse cycles are not synchronized might break down (see Discussion).

\section{Synchronization of prey dynamics}\label{ap:Synchronization}

Using larger densities of predators than what we observed in Scotland
(Fig. E1 below), it is possible to synchronize the two prey species
in our vole-harrier-grouse model, so that they fluctuate with the
same period, the grouse cycle lagging however one year behind that
of the voles (thus we do not really observe synchrony in a strict
ecological sense, but rather ``lagged synchrony''). The typical
6-year period of the grouse cycle (using parameter values of Table
1) is then reduced to the 4-year period of the vole cycle. There is
change in the period of the grouse cycle generated by external forcing,
as is observed for forced oscillators in physics. This phenomenon
is usually referred to as frequency locking or frequency entrainment,
as opposed to ``phase locking'' that refers to perfectly in-phase,
fully synchronous dynamics (Pikovsky and Rosenblum 2007). The dynamical
behaviour observed here is reminiscent of that of simple nonlinear
forced oscillators (e.g. the standard circle map, see Weisstein). 

\textbf{References}

Arkady Pikovsky and Michael Rosenblum (2007) Synchronization. Scholarpedia,
2(12):1459, http://www.scholarpedia.org/article/Synchronization

Weisstein, Eric W. ``Circle Map.'' From MathWorld--A Wolfram Web Resource.
http://mathworld.wolfram.com/CircleMap.html 

\setcounter{figure}{0}

\makeatletter \renewcommand{\thefigure}{E\arabic{figure}} 

\begin{center}
\begin{figure}
\begin{centering}
\includegraphics[width=14cm]{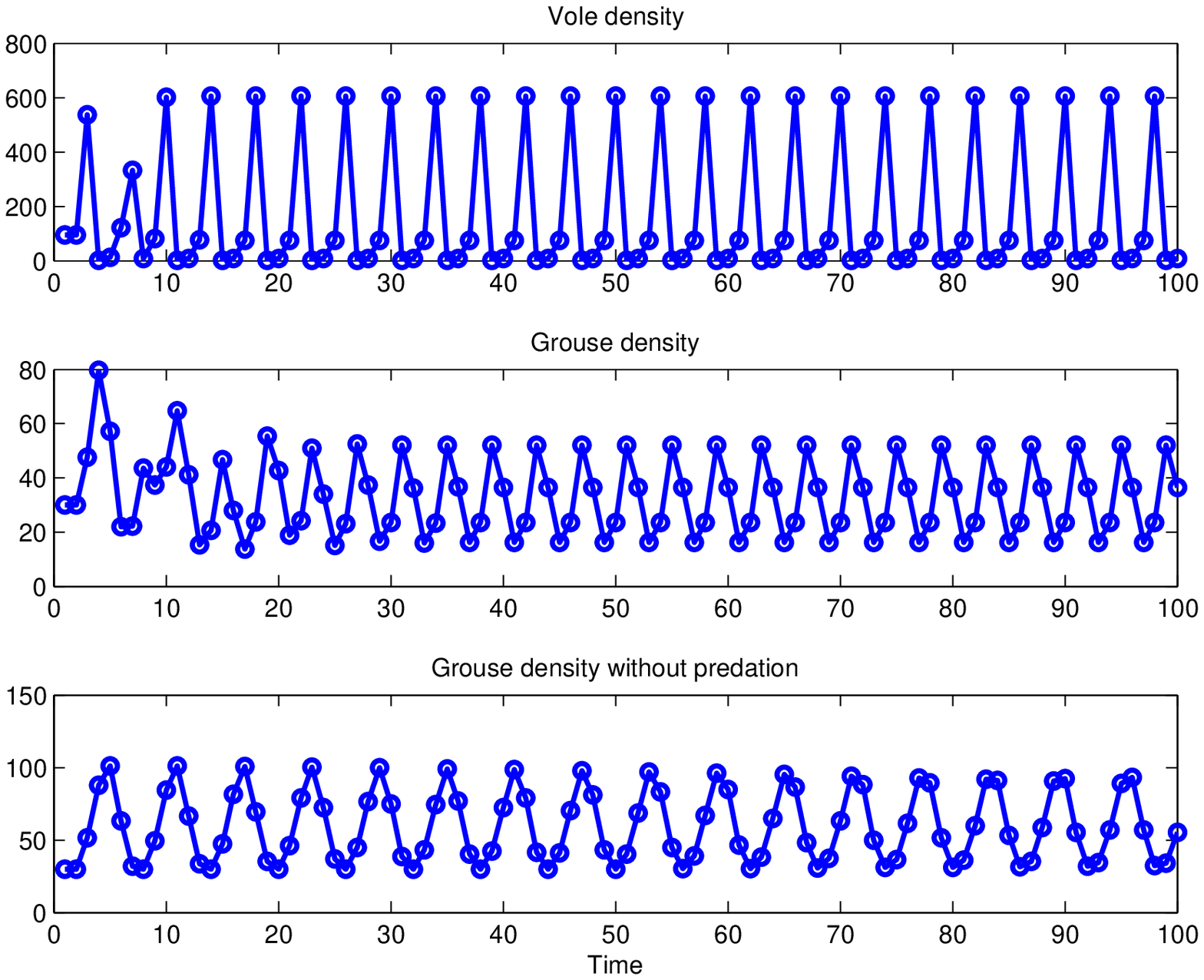}
\par\end{centering}

\caption{Effect of large number of predators (H=0.5 ind.$km^{-2}$) on the
grouse periodicity (middle row), when the system is forced by a strong
vole cycle (top row). The bottom row shows the grouse cycle in the
absence of predation on grouse }

\end{figure}

\par\end{center}

\end{document}